\documentclass[aps,prb,amsmath,amssymb,twocolumn,superscriptaddress,longbibliography,footinbib]{revtex4-2}
\usepackage{graphicx}
\usepackage{tikz}
\usepackage{epstopdf}
\usepackage{mathtools}
\usepackage{bm}
\usepackage{amsmath}
\usepackage{mathrsfs}
\usepackage{xr-hyper}
\usepackage{xcite}
\externalcitedocument{SM}
\usepackage[colorlinks=true,linkcolor=blue,citecolor=blue,urlcolor=blue]{hyperref}
\usepackage{units}
\usepackage{booktabs}
\usepackage{dcolumn}
\usepackage{multirow}
\usepackage{xcolor,colortbl}
\usepackage{dcolumn}
\newcolumntype{d}[1]{D{.}{.}{#1}}

\usepackage{soul}
\usepackage[color=orange!20]{todonotes} 

\usepackage{ulem}
\ifx\nocolor\undefined 
       
   \newcommand{\tred}[1]{\textcolor{red}{#1}} 
   
\else
    
   \newcommand{\tred}[1]{{#1}}
   
\fi

\newcommand{\RNum}[1]{\uppercase\expandafter{\romannumeral #1\relax}}
\usepackage{lipsum}

\newcolumntype{P}[1]{>{\centering\arraybackslash}p{#1}}
\newcolumntype{M}[1]{>{\centering\arraybackslash}m{#1}}

\newcommand{\bfs}{BaFe$_2$S$_3$}	
\newcommand{\bfse}{BaFe$_2$Se$_3$}
\newcommand{\bfx}{BaFe$_2$X$_3$}

\begin{document}

\title{Ground State of {\bfs} from Lattice and Spin Dynamics}

\author{Y. Oubaid}
\thanks{These authors contributed equally to this work.}
\affiliation{Universit\'e Paris-Saclay, CNRS, Laboratoire de Physique des Solides, 91405, Orsay, France.}
\affiliation{Synchrotron SOLEIL, L\'Orme des Merisiers, Saint Aubin BP 48, 91192, Gif-sur-Yvette, France}

\author{S. Deng}
\thanks{These authors contributed equally to this work.}
\affiliation{Institut Laue-Langevin, 71 avenue des Martyrs, 38042 Grenoble, France}

\author{N.S. Dhami}
\affiliation{Universit\'e Paris-Saclay, CNRS, Laboratoire de Physique des Solides, 91405, Orsay, France.}

\author{M. Verseils}
\affiliation{Synchrotron SOLEIL, L\'Orme des Merisiers, Saint Aubin BP 48, 91192, Gif-sur-Yvette, France}
\author{P. Fertey}
\affiliation{Synchrotron SOLEIL, L\'Orme des Merisiers, Saint Aubin BP 48, 91192, Gif-sur-Yvette, France}
\author{P. Steffens}
\affiliation{Institut Laue-Langevin, 71 avenue des Martyrs, 38042 Grenoble, France}

\author{D. Bounoua}
\affiliation{Universit\'e Paris-Saclay, CNRS-CEA, Laboratoire L\'eon Brillouin, 91191, Gif sur Yvette, France}

\author{A. Forget}
\affiliation{Universit\'e Paris-Saclay, CEA, CNRS, SPEC, 91191, Gif-sur-Yvette, France.}

\author{D. Colson}
\affiliation{Universit\'e Paris-Saclay, CEA, CNRS, SPEC, 91191, Gif-sur-Yvette, France.}

\author{P. Foury-Leylekian} \affiliation{Universit\'e Paris-Saclay, CNRS,
  Laboratoire de Physique des Solides, 91405, Orsay, France.}

\author{M.B. Lepetit}
\email[Corresponding author:]{Marie-Bernadette.Lepetit@neel.cnrs.fr}
\affiliation{Institut N\'eel, CNRS, 38042 Grenoble, France}
\affiliation{Institut Laue-Langevin, 71 avenue des Martyrs, 38042 Grenoble, France}

\author{V. Bal\'edent}
\email[Corresponding author:]{victor.baledent@universite-paris-saclay.fr}
\affiliation{Universit\'e Paris-Saclay, CNRS, Laboratoire de Physique des Solides, 91405, Orsay, France.}
\affiliation{Institut universitaire de France (IUF)}

\date{\today}

\begin{abstract}
  We investigate the interplay between lattice symmetry, phonons, and
  magnetism in the quasi-one-dimensional ladder compound BaFe$_2$S$_3$ by
  combining polarized synchrotron infrared spectroscopy, hybrid-functional
  density functional theory calculations, and inelastic neutron
  scattering. Lattice-dynamics analysis reveals that the crystal symmetry is
  lower than previously proposed and is consistent with a $P1$ space group at
  low temperature. Several infrared-active phonon modes exhibit pronounced
  anomalies at both the structural transition temperature
  $T_S \approx 125$--$130$~K and the Néel temperature $T_N \approx
  95$~K. First-principles calculations show that the modes affected at $T_S$
  predominantly involve displacements that modulate magnetic exchange
  pathways. Neutron scattering demonstrates that below $T_N$ the magnetic
  order is three-dimensional, long-ranged, and static. Between $T_N$ and
  $T_S$, the system displays three-dimensional short-range dynamic magnetic
  correlations, which disappear above $T_S$. The structural transition thus
  coincides with the onset of magnetic fluctuations rather than with static
  magnetic order.  Our results point toward the fact that the \bfs\ compound
  could be another example where short-range, dynamical magnetic correlations
  are sufficient to drive a static structural instability, providing a
  magnetically driven mechanism reminiscent of the iron-pnictide 122
  family. These findings highlight the central role of magnetoelastic coupling
  in iron-based superconductors beyond the itinerant regime.
\end{abstract}


\maketitle

\section{Introduction}

Iron-based compounds with strongly correlated electrons and magnetic order
have been an active area of research for the past few decades.  This field,
was ignited by the 2008 discovery of superconductivity at high-temperatures in
the $\rm~LaO_{1-x}F_xFeAs$ compounds ($T_c=\unit[26]{K}$ ~at ambient
pressure\cite{Kamihara2008} and $\unit[43]{K}$ under
$\unit[4]{GPa}$~\cite{Takahashi2008}). Since then $T_c$ above $\unit[100]{K}$
was observed in a monolayer film of $\rm FeSe$ on a doped-$\rm SrTiO_3$
substrate~\cite{FeSe-monolayer-supra}. On top of superconductivity the
iron-based compounds have been shown to display rich phase diagrams with
various magnetic orders~\cite{dai2015antiferromagnetic}, orbital
orders~\cite{patel2019fingerprints,Hosoi2020,Shin2010},
nematicity~\cite{bohmer2022nematicity},
multiferroicity~\cite{Zheng2020,Elbio2014}, etc.
 
Recently, the spin-ladder \bfx\ (X = S, Se) family was shown to exhibit both
quasi-1D-superconductivity under pressure ($T_c$ between $\unit[10]{K}$ and
$\unit[26]{K}$ under
$\unit[9-26]{GPa}$~\cite{Takahashi2015,Yamauchi2015,Ying2017PRB,Birgeneau2019,Pascale2022}), 
multiferroicity~\cite{Zheng2020,Elbio2014}
and irreducible ferrielectricity~\cite{Tian2020}.

\begin{figure}[!t]
    \centering
    \begin{minipage}{.5\linewidth}
    \includegraphics[width=.99\linewidth]{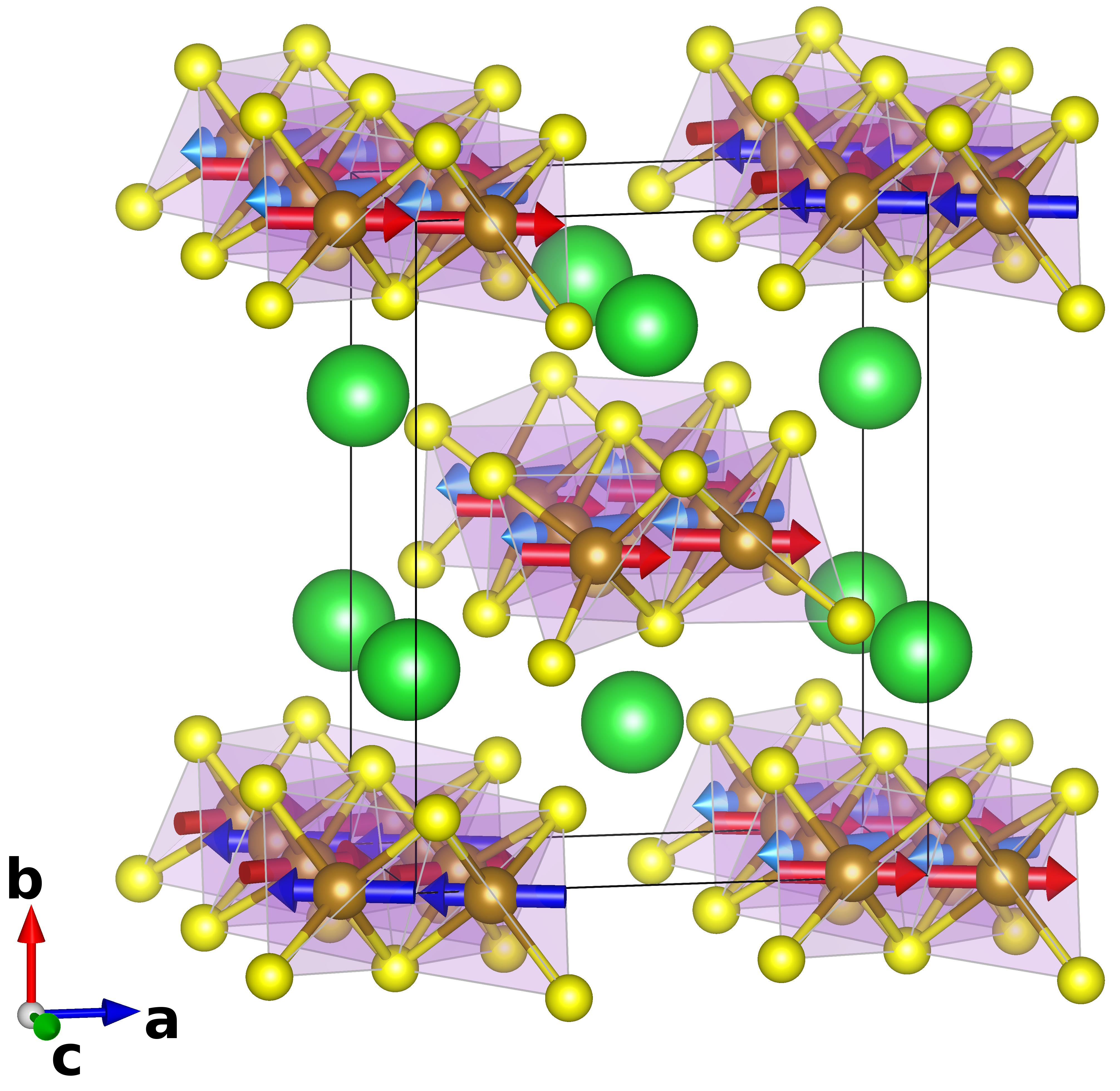}
    \end{minipage}
    \begin{minipage}{.48\linewidth}
    \includegraphics[width=.99\linewidth]{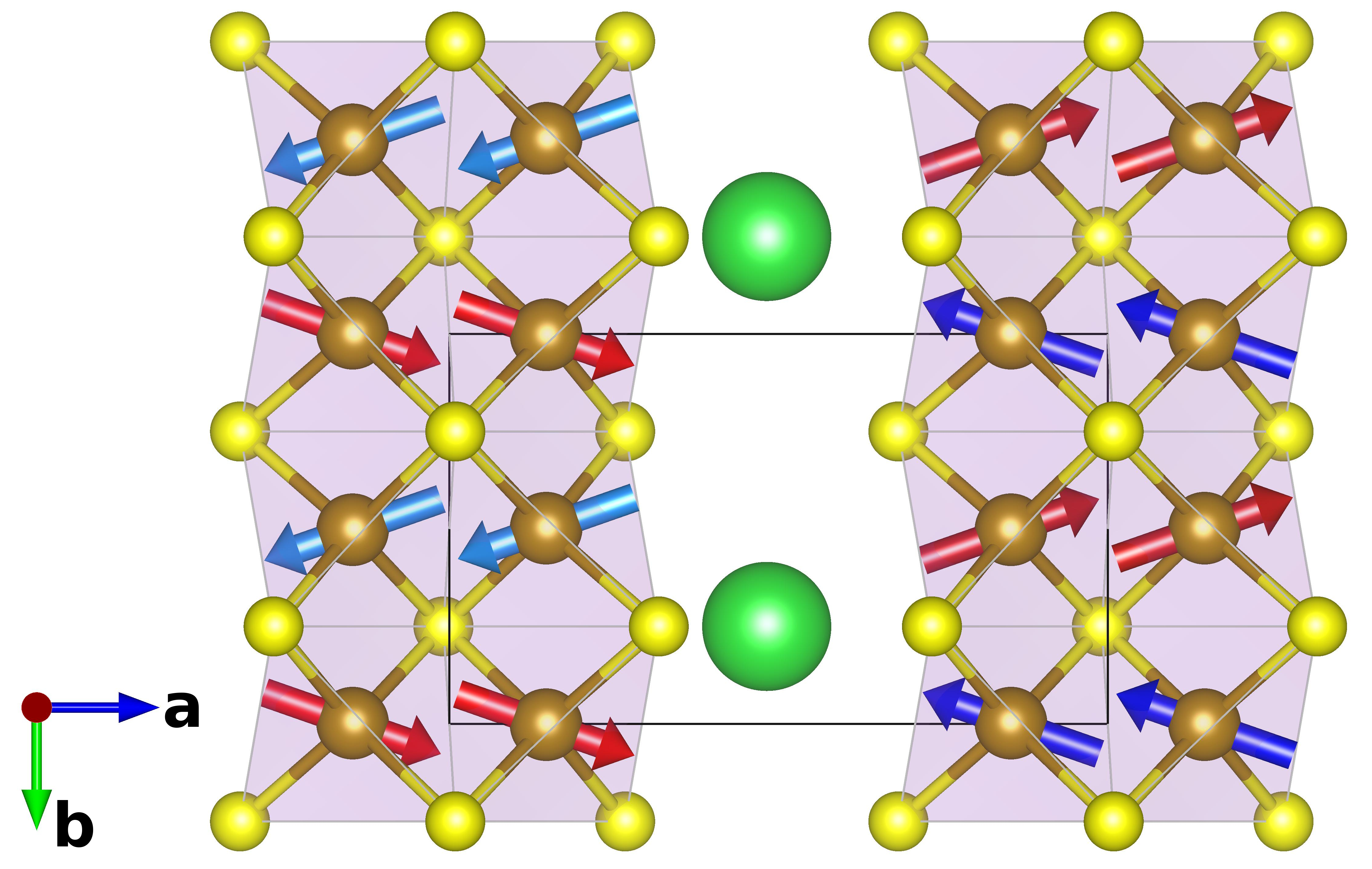}
    \end{minipage}
    \caption{Schematic picture  of the atomic and magnetic structure of {\bfs}. The axes
      follow the $Cmcm$ setting.
      Ba: green, Fe: brown, S: yellow. 
      The arrows show the magnetic moments on different Fe sites.}
    \label{fig:bfs_T_phases}
\end{figure}
\bfx\ compounds are composed of one-dimensional ladders of edge-sharing
Fe$^{2+}$(X$^{2-}$)$_4$ tetrahedra separated by Ba$^{2+}$ ions. While the
\bfse\ compound exhibits an unconventional square-block antiferromagnetic
(AFM) order~\cite{Caron2011,Krzton-Maziopa_2011,Nambu2012}, the
\bfs\ one presents AFM
stripes~\cite{Takahashi2015,Hirata2015,chan2020three_low_Tn} (see
Fig.~\ref{fig:bfs_T_phases}).
In this \bfx\ family, the nature of the insulating ground state at ambient
pressure remains under debate. While some authors report the coexistence of
localized and itinerant electrons~\cite{Mizokawa2015,2017Ohgushi}, suggesting
orbital-selective Mott physics~\cite{patel2019fingerprints}, optical
spectroscopy points alternatively to a magnetically driven Slater
gap~\cite{Jungseek2011}.

Under pressure these correlated insulators turn metallic and then
superconducting~\cite{Takahashi2015,Chi2016,Materne2019,Ying2017PRB,Zheng2018,Birgeneau2019,Pascale2022}.
While the phase of the \bfs\ compound remains unchanged until the
insulator-to-metal transition~\cite{Chi2016,Zheng2018}, the \bfse\ undergoes a
transition from the block magnetic order to the stripe one around
\unit[4]{GPa}~\cite{Ying2017PRB,Birgeneau2019,Pascale2022,Roll2023}.

Both the {\bfs} and {\bfse} compounds had been initially characterized, by
standard X-ray diffraction, as centrosymmetric under ambient
conditions~\cite{HONG1972}.  However, recent pieces of work, high-resolution
X-ray diffraction~\cite{Zheng2020} and combination of theoretical DFT
calculations with Infra-Red (IR) spectroscopy~\cite{Weseloh2022}, revealed
that {\bfse} exhibits weak polar distortions and crystallizes in a
non-centrosymmetric space group even at room temperature. As the system is
cooled down this ferroelectric phase coexists with the magnetically ordered
phase giving rise to multiferroicity.

The \bfs\ compound was first reported to belong to the $Cmcm$ space group over
the whole temperature range~\cite{wu2020iron}. Recently however
high-resolution X-ray experiments reported a structural phase transition at
$T_S=\unit[125-130]{K}$. In addition, weak polar distortions were discovered
over the whole temperature range. The room temperature space group was
established to be $Cm2m$ (let us note that in this setting of the standard
$Amm2$ group, the $Cmcm$ axes are preserved and will be kept all along the
manuscript), and the low temperature one ($T<T_S$) to be
$Pb2_1m$~\cite{Oubaid2026BaFe2S3}.  In this sample the magnetic transition
sets up at $T_N=\unit[95]{K}$~\cite{Oubaid2026BaFe2S3} and does not seem to be
associated with a structural change. A weak resistivity anomaly was also
reported around
$T^*=\unit[180]{K}$~\cite{Yamauchi2015,patel2019fingerprints,Hosoi2020},
attributed by the authors to an orbital ordering, and once more not associated
to structural change.  For sake of completeness one should mention that some
authors also claimed the existence of a spin-glass phase below
$T_G=\unit[25]{K}$~\cite{Gonen2000}, this phase was however never confirmed by
other groups.

As lattice dynamics is well known to be a very thorough test of the symmetry of
a compound, even more effective than high-resolution X-ray diffraction to decipher weak
symmetry breaking, we will conduct in this paper a lattice dynamics study of
\bfs, combining IR measurements and Density
Functional Theory (DFT) phonons calculations.

At this point one should cite that the only available low-temperature lattice
dynamics study of {\bfs} claims the latter to be centrosymmetric with $Cmcm$
space group. This work is based on a combined Raman spectroscopy and
non-magnetic DFT calculation of the phonon modes using the PBE
functional~\cite{2015PRB_CPetrovic}.  As mentioned above, the $Cmcm$ space
group has already been outdated.  Let us point out that the non-magnetic
DFT-PBE calculation exhibits a metallic behavior, while the \bfs\ is a Mott
   insulator~\cite{Mizokawa2015}. In addition it is well known that
the PBE functional has a tendency to over-delocalization. As a consequence one
can expect that the overlook of the magnetic ordering and the use of the PBE
functional will prevent the possibility to decipher weak symmetry breakings
(such as the $Cmcm$ versus $Cm2m$ symmetry breaking experimentally observed at
room temperature in \bfs~\cite{Oubaid2026BaFe2S3}).

\section{Methods}
\subsection{Single Crystal Sample Synthesis} {\bfs} single crystals were grown
from the melt using a self-flux method~\cite{Amigo2021-sub}. Stoichiometric
amounts of BaS (\unit[99.9]{\%}), Fe (\unit[99.9]{\%}), and S
(\unit[99.999]{\%}) powders were mixed in a BaS:Fe:S molar ratio of
1:2.05:3. Approximately \unit[2]{g} of the mixture were pelletized, placed in a carbon
crucible, and sealed in an evacuated quartz ampoule backfilled with a partial
pressure of \unit[300]{mbar} of argon.

The ampoule was heated in a vertical tubular furnace to \unit[1100]{$^\circ$C}
and held at this temperature for \unit[24]{h} to ensure homogenization. It was then
slowly cooled to \unit[750]{$^{\circ}$C} at a rate of
\unit[6]{$^{\circ}$C/h}, followed by furnace cooling to room
temperature. This procedure yielded a centimeter-sized pellet composed of
densely packed, millimeter-sized rod-shaped crystals naturally co-aligned
along the $\vec{c}$ axis of the orthorhombic $Cmcm$ structure (ladder-leg
direction).

The crystalline quality and bulk properties of these samples were thoroughly
characterized in a previous study, confirming their high structural and
chemical homogeneity~\cite{Oubaid2026BaFe2S3}.

\subsection{Polarized Infrared Measurements}
Millimeter-sized single crystals were selected and pre-oriented using X-ray
diffraction at the Laboratoire de Physique des Solides. The crystals were
mounted on a copper sample holder using silver paste to ensure optimal thermal
contact. Particular care was taken to orient the samples such that the
incident electric and magnetic field components were well defined with respect
to the three crystallographic axes of the $Cmcm$ structure.

Infrared reflectivity measurements were performed at the AILES beamline of the
SOLEIL synchrotron using a Bruker IFS125 Michelson interferometer. The
experimental setup included a closed-cycle helium cryostat, a \unit[4.2]{K}
bolometer detector, and a \unit[6]{$\mu$m} beamsplitter, providing a spectral
resolution of \unit[3]{cm$^{-1}$}. The incident radiation was linearly polarized using
a polyethylene polarizer. Reflectivity spectra were collected in the
temperature range \unit[20–300]{K}. Absolute reflectivity was determined using an in
situ gold evaporation technique, enabling accurate reference measurements.

\subsection{Inelastic Neutron Scattering}
Inelastic neutron scattering experiments were carried out on the cold-neutron
triple-axis spectrometer THALES at the Institut Laue-Langevin
(ILL)~\cite{BFS2025_ILL}. Pyrolytic graphite (PG) monochromator and analyzer
were used. The sample temperature was controlled between \unit[150]{K} and
\unit[10]{K} using a standard orange helium-flow cryostat.

The final neutron wave vector was fixed to \unit[$k_f=$1.5]{~\AA$^{-1}$},
providing an optimal compromise between neutron flux and energy width of the
Bragg peak (better than \unit[50]{$\mu$eV}). The data were indexed in the
orthorhombic $Cmcm$ setting with lattice parameters \unit[$a=$8.75]{\AA} (rung
direction), \unit[$b =$11.19]{\AA} (perpendicular to the ladder plane), and
\unit[$c=$5.28]{\AA} (ladder-leg direction).

A single-crystal of approximately \unit[0.5]{cm$^{3}$} was aligned in the
$(H,H,L)$ scattering plane defined by the reciprocal lattice vectors $(H,H,0)$
and $(0,0,L)$. Owing to the reduced value of the neutron wave vector, only a
limited number of magnetic Bragg reflections were accessible; all measurements
were therefore performed at the $(\nicefrac{1}{2},\nicefrac{1}{2},1)$
position. The magnetic Bragg peaks exhibit a mosaic spread comparable to that
of the nuclear $(1,1,0)$ and $(2,0,0)$ reflections.

\subsection{Theoretical Calculations}
The $ab-initio$ calculations were performed using DFT.  In order to best
account for self-interaction and weak lattice distortions, we used the hybrid
B3LYP functional~\cite{becke1993density} and a localized basis set.  An
all-electrons atomic gaussian basis set of $3\zeta + P$ quality was used for
the Fe and S atoms~\cite{vilela2019bsse}.  The Ba atoms were represented using
a relativistic core pseudo-potential of the Stuttgard group
(ECP46MWB~\cite{kaupp1991pseudopotential}) and the associated basis set where
the diffuse functions exponents were taken to be 1.2 as recommended by
Scuseria $et.\ al$~\cite{heyd2005energy}.  The calculations were done using
the CRYSTAL23 code~\cite{erba2022crystal23}.  The shrinking factor was set so
that to sample the Brillouin zone with an approximate grid of
\unit[0.016$\pm$0.002]{\AA$^{-1}$} intervals in each direction.  The
stripe-like spin order found in neutron scattering
experiments~\cite{Hirata2015,chan2020three_low_Tn} was used in all
calculations, geometry optimization as well as lattice dynamics.

\section{Results}
\subsection{Lattice Dynamics}
\subsubsection{Polarized Infrared spectroscopy}
\begin{figure*}[t]
  \centering

  \begin{tabular*}{\textwidth}{@{\extracolsep{\fill}}ccc}
    \multicolumn{1}{c}{\small $\vec E\!\parallel\!\vec a$ (\textbf{Ladder-rung})} &
    \multicolumn{1}{c}{\small $\vec E\!\parallel\!\vec b$ (\textbf{Perpendicular})} &
    \multicolumn{1}{c}{\small $\vec E\!\parallel\!\vec c$ (\textbf{Ladder-leg})} \\[2pt]

    \includegraphics[width=0.32\textwidth]{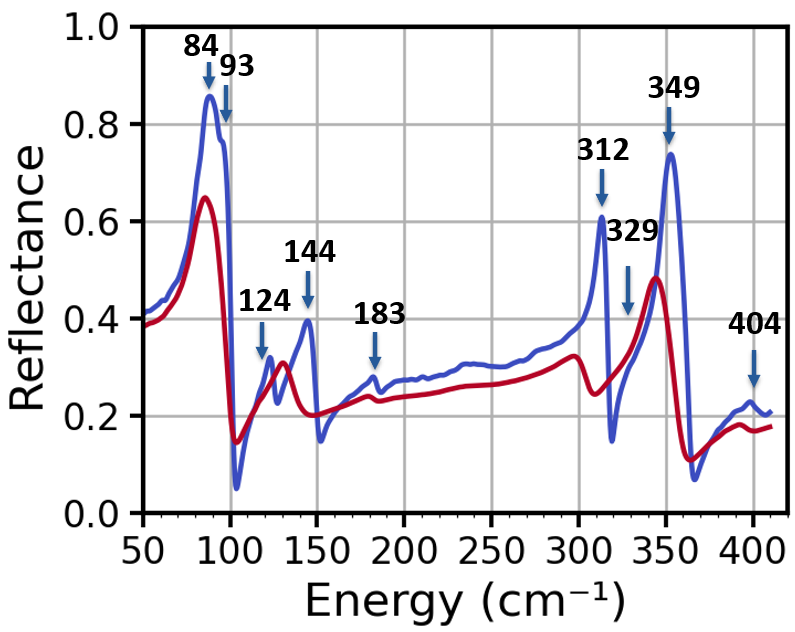} &
    \includegraphics[width=0.32\textwidth]{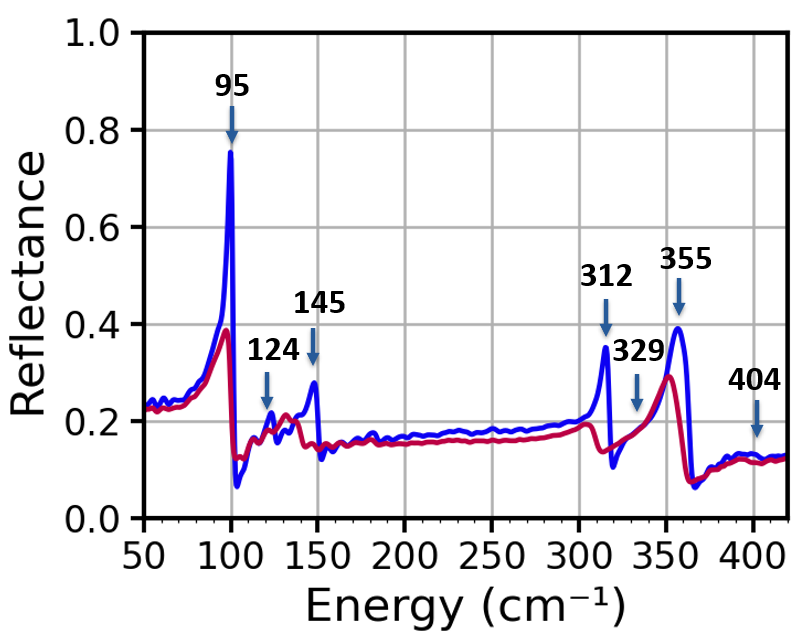} &
    \includegraphics[width=0.32\textwidth]{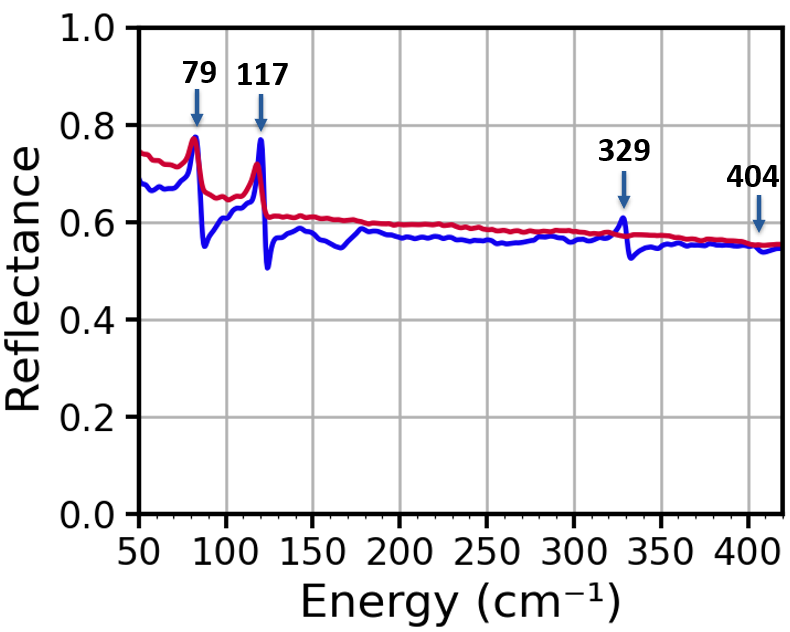} \\[2pt]

    \includegraphics[width=0.32\textwidth]{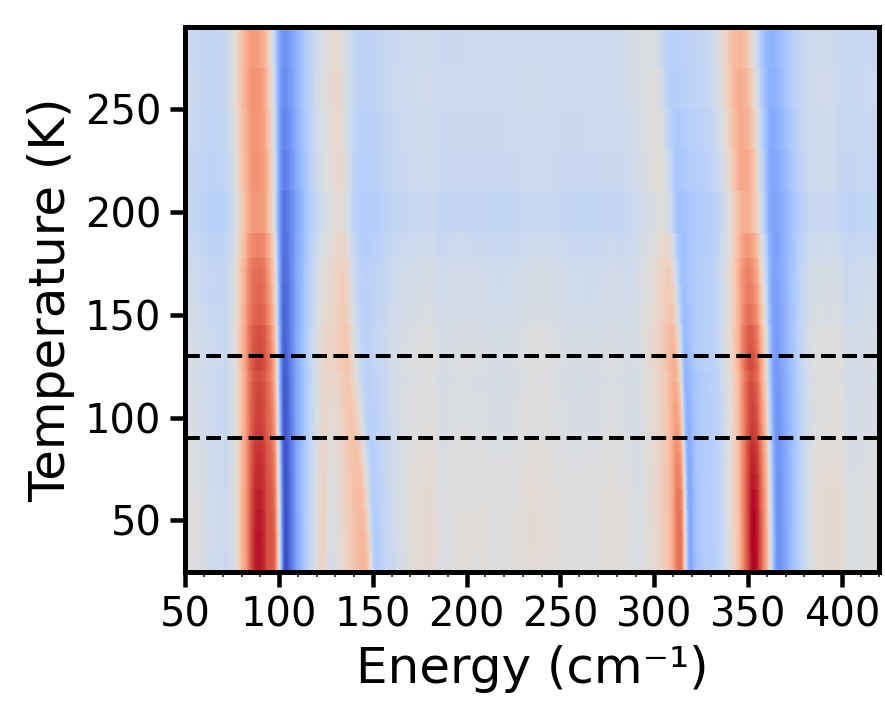} &
    \includegraphics[width=0.32\textwidth]{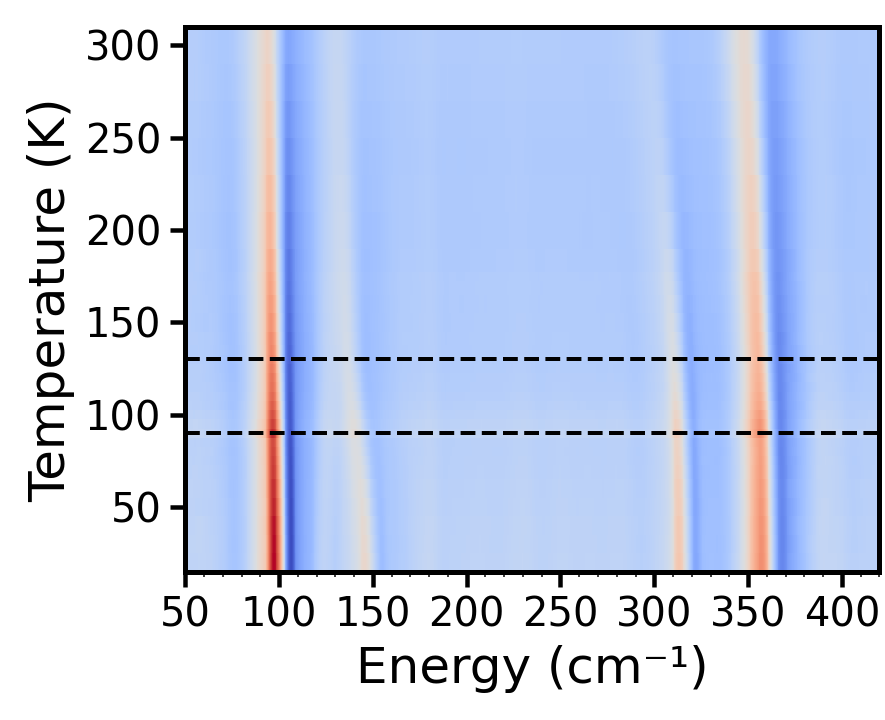} &
    \includegraphics[width=0.32\textwidth]{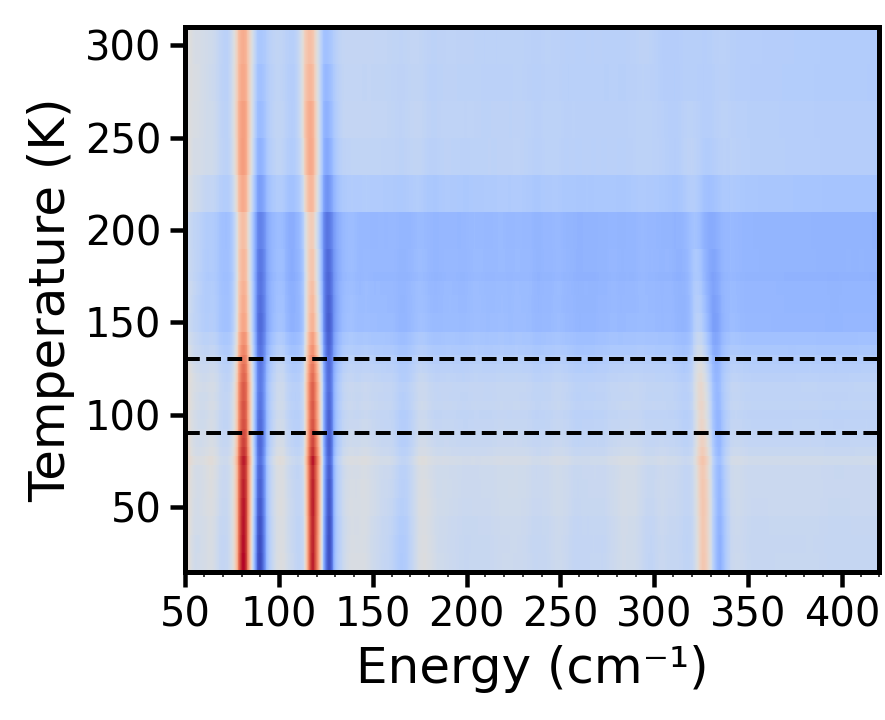} \\[2pt]

    \multicolumn{3}{c}{\includegraphics[width=0.25\textwidth]{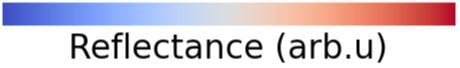}} \\
  \end{tabular*}

  \caption{ \textbf{Temperature-dependent reflectance spectra.}  (Top row)
    Reflectance measured at room temperature (red) and at the lowest
    accessible temperature \unit[20]{K} (blue), with the electric field
    $\vec E$ applied along the three crystallographic directions of the $Cmcm$
    space group.  The arrows mark the fitted IR-active phonon modes at
    \unit[20]{K}.  
    (Bottom row) Heat maps showing the evolution of the
    reflectance as a function of temperature for the three directions, after
    background subtraction.  Horizontal black dashed lines, marking $T_{N}$,
    and $T_{S}$ in ascending order, are added to guide the eye.  The
    reflectance intensity (in arbitrary unit) scales from low (blue) to high
    (red).  }
  \label{fig:Reflectance_all_directions}
\end{figure*}

We measured the reflectance in the far-infrared range, with the electric field
$\vec E$ applied in all three $Cmcm$ lattice directions.  Let us note that in
this paper we will always refer to the $Cmcm$ axes, where $\vec a$ is the
ladder-rung direction, $\vec c$ is the ladder-leg direction and $\vec b$ is
perpendicular to the latters.
  
Fig.~\ref{fig:Reflectance_all_directions} shows the reflectance spectra
measured at \unit[300]{K} (red) and \unit[20]{K} (blue), together with the
corresponding heat maps illustrating the evolution of the phonon energies and
intensities as a function of temperature.  We used the Drude-Lorentz model to
analyze the dielectric function.  The details of the experimental fit can be
found in the Supplementary Information~\cite{SI} (SI) including Ref.~\cite{kuzmenko2005kramers}. 

\subsubsection{First-principle Calculations \& Symmetry Analysis}
We first performed the $\Gamma$-point lattice dynamic calculations, after full
geometry optimization, for the two space groups cited in the literature for the
low temperature phase, namely $Cmcm$~\cite{HONG1972} and
$Pb2_1m$~\cite{Oubaid2026BaFe2S3}. The total energy of the system was found to
be \unit[2.79]{eV/f.u.} lower for the $Pb2_1m$ group, in agreement with the
recent X-ray results of Ref.~\onlinecite{Oubaid2026BaFe2S3}.

Lattice dynamics calculations, at the $\Gamma$ point, under harmonic
approximation, indicated that both phases can be dynamically stable, without
any imaginary phonon modes. Symmetry analysis yields the following
irreducible representations (irrep) for the optical modes
{\small
\begin{eqnarray*} 
  Cmcm &:& \underbrace{\overbrace{4B_{1u}}^{\vec E \parallel \vec c}
           \oplus \overbrace{5B_{2u}}^{\vec E \parallel \vec b}
           \oplus \overbrace{4B_{3u}}^{\vec E \parallel \vec a}}_{\text{IR active}}
           \oplus
           \underbrace{5A_g \oplus 6B_{1g} \oplus 3B_{2g} \oplus 4B_{3g}}_{\text{Raman active}} \\ 
  Pb2_1m&:& \underbrace{\overbrace{21A_1}^{\vec E \parallel \vec b}
            \oplus \overbrace{13 B_1}^{\vec E \parallel \vec c}
            \oplus \overbrace{21 B_2}^{\vec E \parallel \vec a}}_{\text{IR+Raman active}}
            \underbrace{\oplus 14 A_2}_{\text{Raman active}}
\end{eqnarray*}
} Again the symmetry analysis discards the $Cmcm$ space group, as for instance it yields only
$4B_{3u}$ modes  when the light polarization is parallel to $\vec a$,
while our IR experiment clearly sees 9 (see Fig.~\ref{fig:Reflectance_all_directions}). 

Table~\ref{tab:DFT_Phonons} displays the DFT phonons frequencies for the
$Pb2_1m$ group and the best fit to the experimental values taking into account
the light polarization. Several modes seen in the experiment cannot be
assigned in the $Pb2_1m$ group. In both the $B_2$ and $A_1$ irreps, the
\unit[312]{cm$^{-1}$} and \unit[329]{cm$^{-1}$} modes, seen in both polarizations,
cannot be simultaneously assigned. In the $B_1$ irrep, the
\unit[329]{cm$^{-1}$} and \unit[404]{cm$^{-1}$} modes do not have DFT counterparts.

\begin{table}[h!]
  \caption{DFT frequencies (cm$^{-1}$) of optical phonons in the $Pb2_1m$
    space group and assignment of the IR modes measured at \unit[20]{K}. The
    experimental modes typeset in red cannot be assigned to the computed ones,
    either because there is only one theoretical mode for two experimental
    ones, or because there is no theoretical mode computed in the energy
    range and the correct irrep.}
  \label{tab:DFT_Phonons}
\centering
\begin{minipage}[t]{0.3\linewidth}
    \vspace{0pt}
    \begin{tabular}{cc}
      \hline \hline
      DFT   & IR \\
      $B_2$ & $\vec E\parallel \vec a$ \\
      \hline
      47.5  &  \\
      51.5  &  \\
      71.4  &  \\
      78.4  &  \\
      89.5  &  84 \\
      93.8  &  93 \\
      122.8 & 124 \\
      130.5 &     \\
      137.8 &     \\
      139.9 & 144 \\
      182.5 & 183 \\
      198.9 &     \\
      204.2 &     \\
      250.4 &     \\ 
      269.9 &     \\
      272.9 &     \\
      280.5 &     \\
      \multirow{2}*{320.4}  & \tred{\bf 312} \\
            & \tred{\bf 329} \\
      354.4 & 349 \\
      378.5 &     \\
      402.9 & 404 \\
      \hline \hline
     \end{tabular}
   \end{minipage}
   \begin{minipage}[t]{0.3\linewidth}
    \vspace{0pt}
    \begin{tabular}{cc}
      \hline \hline
      DFT   & IR   \\
      $A_1$ & $\vec E\parallel \vec b$ \\
      \hline
      21.1  &   \\
      54.5  &   \\
      57.0  &   \\
      71.9  &   \\
      81.0  &   \\
      85.3  & 95 \\
      116.0 & 124 \\
      133.1 &    \\
      135.4 &    \\
      147.3 & 145 \\
      169.2 &    \\
      192.7 &    \\
      213.5 &    \\
      245.4 &    \\
      276.1 &    \\
      276.3 &    \\
      283.8 &    \\
      \multirow{2}*{324.5} & \tred{\bf 312} \\
            & \tred{\bf 329} \\
      346.2 & 355 \\
      365.2 &    \\
      379.4 &    \\
            & \tred{\bf 404} \\
      \hline \hline
     \end{tabular}
   \end{minipage} 
   \begin{minipage}[t]{0.3\linewidth}
    \vspace{0pt}
    \begin{tabular}{cc}
      \hline \hline
      DFT   & IR   \\
      $B_1$ & $\vec E\parallel \vec c$ \\
      \hline
      39.7  &    \\
      44.7  &    \\
      82.7  & 79 \\
      91.9  &    \\
      100.5 &    \\
      122.0 & 117 \\
      130.2 &     \\
      230.2 &     \\
      241.8 &     \\
      250.9 &     \\
      266.2 &     \\
      283.0 &     \\
      287.1 &     \\
            & \tred{\bf 329} \\
            & \tred{\bf 404} \\           
      \hline \hline
     \end{tabular}
   \end{minipage}
 \end{table}

 The fact that several modes are seen with different light polarization claims
 for a symmetry lowering.  We thus investigated the lattice dynamics for all
 maximal subgroups of $Pb2_1m$, i.e. $P2_1$,  $Pm$ and $Pc$. 
 Symmetry analysis of the irrep conjunction tells us that
 \begin{align*}
   \label{eq:irreps}
   Pm &:& A' &=& A_1 \oplus B_2 &= \left(A_g \oplus B_{2u}\right)  \oplus \left(B_{1g} \oplus B_{3u}\right) \\
      & & A'' &=& B_1 \oplus A_2 &= \left(B_{3g} \oplus B_{1u}\right)  \oplus \left(B_{2g} \oplus A_{u}\right) \\[2ex]
   Pc &:& A' &=& A_1 \oplus B_1 &= \left(A_g \oplus B_{2u}\right)  \oplus \left(B_{3g} \oplus B_{1u}\right) \\
      & & A'' &=& A_2 \oplus B_2 &= \left(B_{2g} \oplus A_{u}\right)  \oplus  \left(B_{1g} \oplus B_{3u}\right) \\[2ex]
   P{2_1} &:& A&=& A_1 \oplus A_2 &= \left(A_g \oplus B_{2u}\right)  \oplus \left(B_{2g} \oplus A_{u}\right) \\
      & & B &=& B_1 \oplus B_2 &=  \left(B_{3g} \oplus B_{1u}\right)  \oplus  \left(B_{1g} \oplus B_{3u}\right) 
 \end{align*}
 At this point one shall remember that in $Cmcm$  the light polarization along the
 $\vec a$ axis addresses phonons in the $B_{3u}$ irrep, the
 polarisarion along $\vec b$ addresses phonons in the $B_{2u}$ irrep and along
 $\vec c$ addresses phonons in the $B_{1u}$ irrep.

 Our calculations in those
 three groups yield similar formation energies as in the $Pb2_1m$ group, thus
 not allowing to discriminate between them.  This is consistent with the fact
 that diffraction experiment was not able to determine the exact structure.
 Going now to the lattice dynamics, unfortunately none of the three subgroups
 could account for all the observed phonons modes when placed in the proper
 irreps (see SI~\cite{SI} for detailed tables). 
 
 We thus finally performed phonon calculation within the $P1$ space group. The
 results are displayed in Table~\ref{tab:DFT_Phonons_P1}. In order to better
 assign the phonons we looked whether the main atomic displacements are
 compatible with the light polarization direction. This time all modes
 can easily be accounted for. The lattice dynamics analysis thus points toward
 the loss of all point group symmetry operations. 
 
 \begin{table}[h!]
   \caption{DFT frequencies (cm$^{-1}$) of optical phonons computed in $P1$ and their
     best experimental assignment. 
     }
  \label{tab:DFT_Phonons_P1}
\centering
\begin{tabular}{d{3.1}rrr  @{\hspace{0.4cm}} d{3.1}rrr}
      \hline \hline
      \multicolumn{1}{c}{DFT}   & \multicolumn{3}{c}{IR} & \multicolumn{1}{c}{DFT}   & \multicolumn{3}{c}{IR} \\
      A   & {\small $\vec E\!\parallel\! \vec a$}
          & {\small $\vec E\!\parallel\! \vec b$}
          & {\small $\vec E\!\parallel\! \vec c$}
    & A   & {\small $\vec E\!\parallel\! \vec a$}
          & {\small $\vec E\!\parallel\! \vec b$}
          & {\small $\vec E\!\parallel\! \vec c$} \\
      \hline
      72.1  &     &     &   &        136.8 &     &     &   \\   
      75.2  & 84  &     &   &        136.8 &     &     &   \\   
      81.8  &     &     &   &        144.1 & 144 & 145 &   \\   
      84.9  &     & 95  &   &        151.6 &     &     &   \\   
      87.5  &     &     &   &        156.9 &     &     &   \\   
      93.3  &     &     & 79&        168.3 &     &     &   \\   
      94.8  &     &     &   &        170.7 & 183 &     &   \\   
      96.2  &     &     &   &        199.7 &     &     &   \\   
      97.1  & 93  &     &   &        200.0 &     &     &      \\
      98.0  &     &     &   &        202.7 &     &     &      \\
      102.8 &     &     &   &        ......&     &     &      \\
      109.1 &     &     &   &        296.7 &     &     &      \\
      110.0 &     &     &117&        318.5 & 312 & 312 &      \\
      113.8 &     &     &   &        336.8 & 329 & 329 & 329  \\
      117.3 &     &     &   &        346.4 & 349 &     &      \\
      121.5 & 124 &     &   &        347.3 &     & 355 &      \\
      122.4 &     & 124 &   &        378.9 &     &     &      \\
      123.0 &     &     &   &        381.3 &     &     &      \\
      135.8 &     &     &   &        383.2 &     &     &      \\
      136.5 &     &     &   &        396.5 & 404 & 404 & 404  \\
      \hline \hline
     \end{tabular}
 \end{table}

 As a summary after easily discarding $Cmcm$, we tested the $Pb2_1m$ group
 found in X-Ray scattering~\cite{Oubaid2026BaFe2S3} as well as all its
 subgroups against the experimental IR data. The experimental to theoretical
 comparison lead us to discard all $k$-index=1 subgroups of $Pb2_1m$, but
 $P1$.  Table~\ref{tab:sum_sym} summarizes these results.

 \begin{table}[h!]
   \caption{Summary of symmetry analysis for all $Pb2_1m$ maximal subgroups. Discard
       reasons for subgroups due to the comparison between experimental and
       computed phonons.}
     \centering
     \begin{tabular}{cc|c|l}
       \hline       \hline
      \multicolumn{2}{c|}{$Pb2_1m$} & \multicolumn{1}{c}{} 
      & \multirow{2}{*}{Arguments to discard $Pb2_1m$} \\
      \cline{1-2}
      \multicolumn{1}{c|}{Irrep} & $\vec E\!\parallel$ & \multicolumn{1}{c}{} &  \\
       \hline
       \multicolumn{1}{c|}{$A_1$} & $\vec a$ & \multicolumn{1}{c}{} & Exp. 312 \& 329 modes seen in  $\vec E\!\parallel\!\vec a\&\vec b$ cannot\\
       \multicolumn{1}{c|}{$B_2$} & $\vec c$ & \multicolumn{1}{c}{} & be assigned simultaneously.  Exp. 404 modes \\
       \multicolumn{1}{c|}{$A_2$} &          & \multicolumn{1}{c}{} &  seen in $\vec E\!\parallel\!\vec b\&\vec c$, exp. 329 mode  seen in  $\vec E\!\parallel\!\vec c$   \\
       \multicolumn{1}{c|}{$B_1$} & $\vec b$ & \multicolumn{1}{c}{} &cannot  be assigned.\\
       \hline     \\[-1ex]  \hline 
       \multicolumn{2}{c|}{$Pb2_1m$} &$Pm$ 
       & \multirow{2}{*}{Arguments to discard subgroup $Pm$} \\
       \cline{1-3}
      \multicolumn{1}{c|}{Irrep} & $\vec E\!\parallel$ &  Irrep &  \\
       \hline
       \multicolumn{1}{c|}{$A_1$} & $\vec a$ & \multirow{2}{*}{$A'$} & \\
       \multicolumn{1}{c|}{$B_2$} & $\vec c$ &                       &  \\
       \hline 
       \multicolumn{1}{c|}{$A_2$} &          &  \multirow{2}{*}{$A"$} & \multirow{2}{*}{Exp. 329 \& 404 modes  cannot be assigned.} \\
       \multicolumn{1}{c|}{$B_1$} & $\vec b$ &                        & \\
       \hline     \\[-1ex]  \hline 
       \multicolumn{2}{c|}{$Pb2_1m$} & $Pc$  
       & \multirow{2}{*}{Arguments to discard subgroup $Pc$} \\
        \cline{1-3} 
      \multicolumn{1}{c|}{Irrep} & $\vec E\!\parallel$ &  Irrep &  \\
       \hline 
       \multicolumn{1}{c|}{$A_1$} & $\vec a$ & \multirow{2}{*}{$A'$} &Exp. 329 \& 349 modes seen in  $\vec E\!\parallel\!\vec a$ cannot   \\
       \multicolumn{1}{c|}{$B_1$} & $\vec b$ &                       & be assigned simultaneously. Idem for the 329  \\
       \cline{1-3} 
       \multicolumn{1}{c|}{$A_2$} &          &  \multirow{2}{*}{$A"$} &  \& 355 seen in  $\vec E\!\parallel\!\vec b$. Exp. 404 modes seen in   \\
       \multicolumn{1}{c|}{$B_2$} & $\vec c$ &                       & $\vec E\!\parallel\!\vec a \& \vec b$ cannot be assigned.\\
       \hline   \\[-1ex]     \hline
       \multicolumn{2}{c|}{$Pb2_1m$} & $P2_1$  
       & \multirow{2}{*}{Arguments to discard subgroup $P2_1$} \\
        \cline{1-3} 
       \multicolumn{1}{c|}{Irrep} & $\vec E\!\parallel$ & Irrep &  \\
       \hline 
       \multicolumn{1}{c|}{$A_1$} & $\vec a$ & \multirow{2}{*}{$A$} &Exp. 329 \& 349 modes seen in  $\vec E\!\parallel\!\vec a$ cannot    \\
       \multicolumn{1}{c|}{$A_2$} &  &                       &be assigned simultaneously. Idem for the 329 \\
       \cline{1-3} 
       \multicolumn{1}{c|}{$B_1$} & $\vec b$ &  \multirow{2}{*}{$B$} & \& 355 seen in  $\vec E\!\parallel\!\vec b$.  Exp. 404 modes seen in   \\
       \multicolumn{1}{c|}{$B_2$} & $\vec c$ &                       &$\vec E\!\parallel\!\vec c$ cannot be assigned. \\
      \hline       \hline
     \end{tabular}
     \label{tab:sum_sym}
   \end{table}

\subsubsection{Evolution of phonons with respect to temperature}
\begin{figure*}[t]
    \centering
    \includegraphics[width=0.8\linewidth]{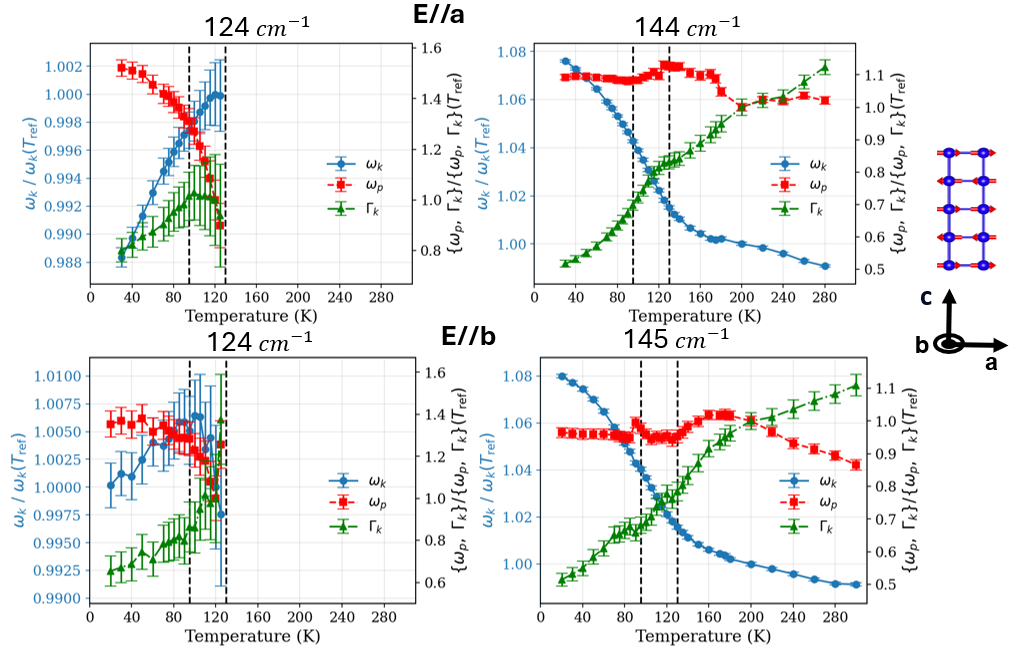}
  \hspace*{-4.5ex}  \includegraphics[width=0.735\linewidth]{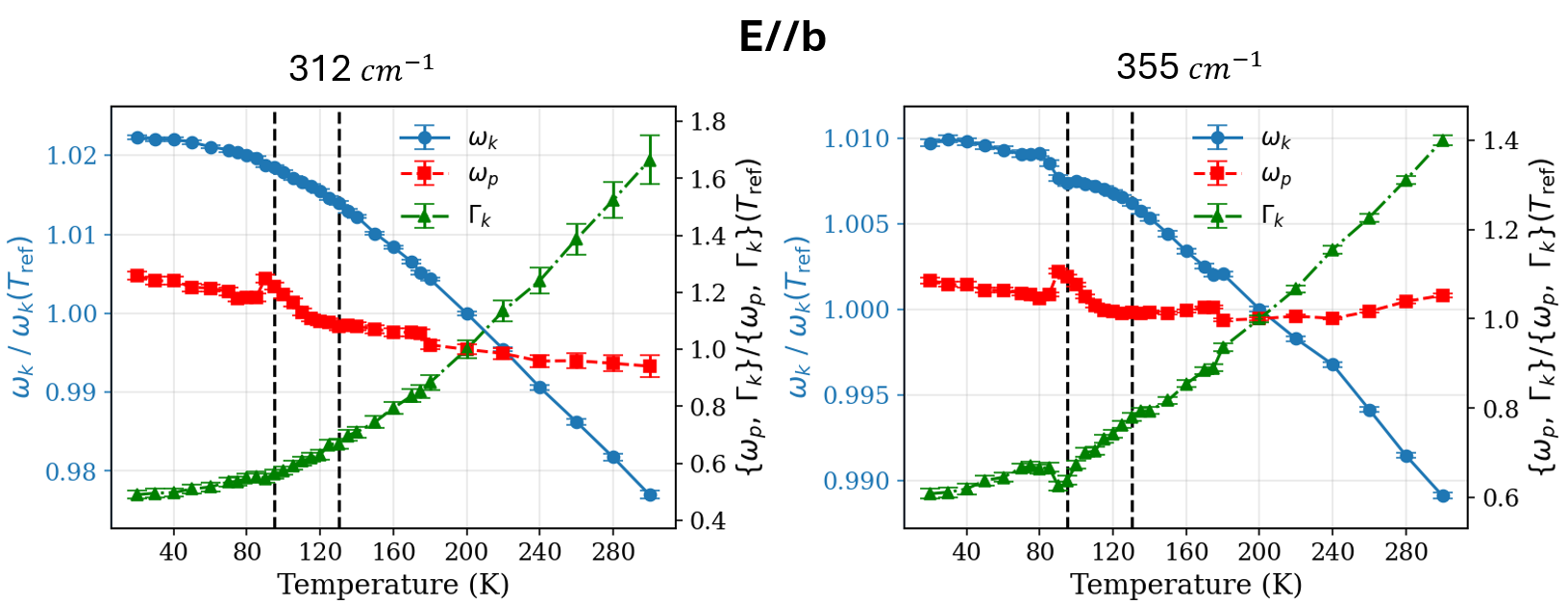} 
  \caption{Temperature evolution of the \unit[124]{cm$^{-1}$},
    \unit[145]{cm$^{-1}$}, \unit[312]{cm$^{-1}$} and \unit[355]{cm$^{-1}$}
    phonons modes. The two \unit[124]{cm$^{-1}$} modes disappear at
    $T_S$. Blue curves~: oscillator frequencies, $\omega_k$.  Red curves~:
    normalized plasma frequencies, $\omega_p$.  Green curves~: width,
    $\Gamma_k$. }
    \label{fig:Tdep}
\end{figure*}

Looking at the temperature dependence of the phonon modes, one can see that
several phonons (at \unit[145]{cm$^{-1}$}, \unit[312]{cm$^{-1}$} and
\unit[355]{cm$^{-1}$}) exhibit anomalies in their intensities at the magnetic
transition (see Fig.~\ref{fig:Tdep}).  This observation clearly points toward
the existence of magneto-elastic coupling.  These phonons are
seen when the light polarization is either along the $\vec a$ or $\vec b$. 

Above the structural transition, $T_S$, three phonons disappear i.e. the
phonon seen at \unit[93]{cm$^{-1}$} (see SI Fig.~\ref{SM_fig:93cm}) and
\unit[124]{cm$^{-1}$} when the light polarization is along $\vec a$ (ladder
rungs) and the phonon seen at \unit[124]{cm$^{-1}$} when the light
polarization is along the $\vec b$ direction. In addition, the phonon at
\unit[145]{cm$^{-1}$} exhibit anomalies at $T_S$ but its intensity does not
dies out above $T_S$.

Let us look now at the displacement vectors of the two modes at
\unit[124]{cm$^{-1}$} (see SI Fig.~\ref{SM_fig:Anomaly_Ts_DFT_124}). 
The first mode seen when $\vec E\!\parallel\!\vec a$  
mainly involves Fe movements along the ladder rungs. 
The second one, seen when $\vec E\!\parallel\!\vec b$, 
mainly involves movements of the sulfur atoms along the ladder legs.

\begin{figure}[htbp]
    \centering
    \includegraphics[width=0.98\linewidth]{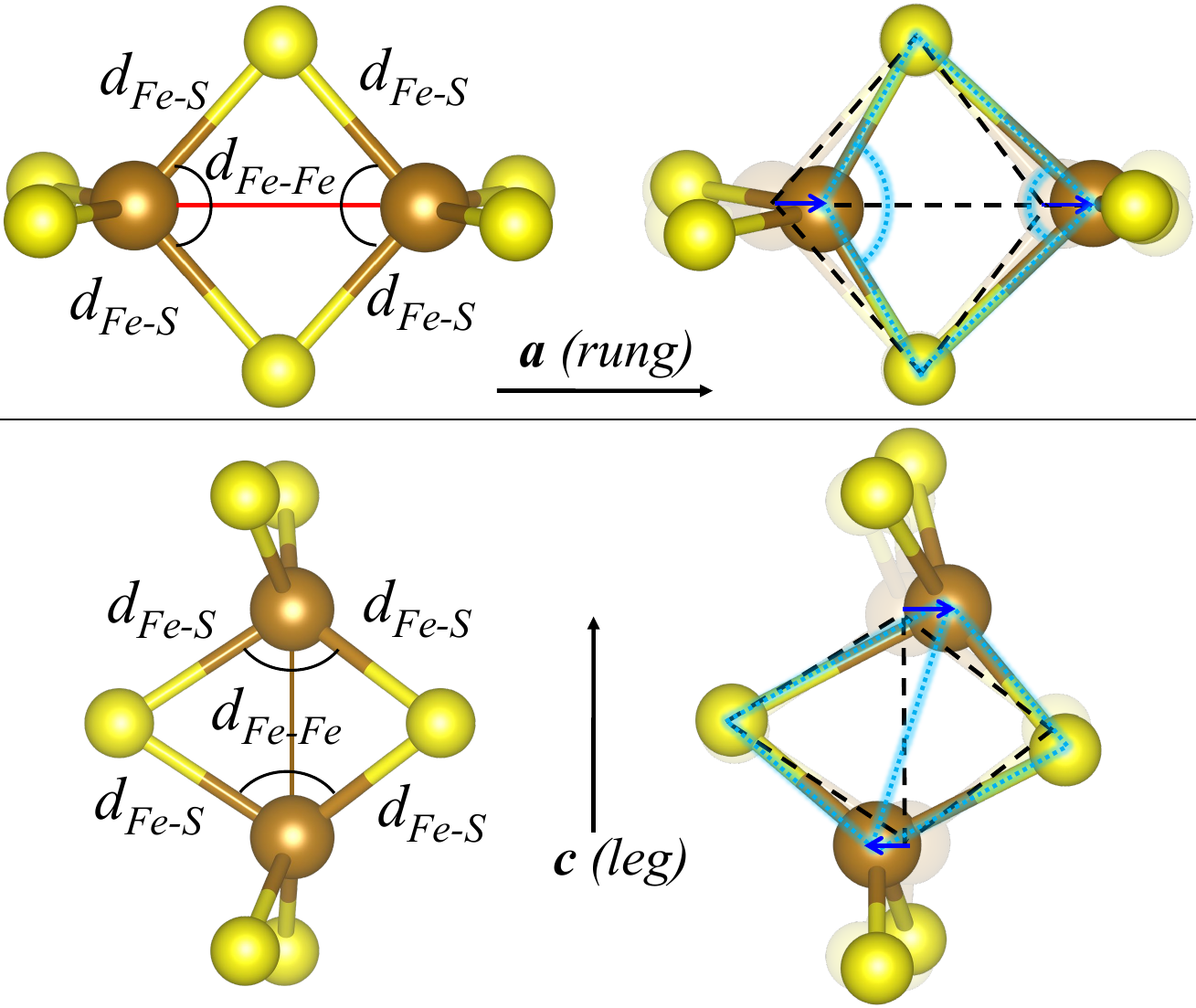}
  \caption{
  Schematic representation of the magnetic exchange interactions 
  along the ladder rung ($J_{\text{rung}}$) and leg ($J_{\text{leg}}$) directions.  
  The equilibrium position is shown on the left, in comparison with the 
  atomic displacements of the calculated phonon mode 
  at \unit[121.5]{cm$^{-1}$}, 
  assigned to the mode at \unit[124]{cm$^{-1}$} 
  seen when $\vec E\!\parallel\!\vec a$. 
  The main effects are highlighted by blue in both cases, respectively.
  }
    \label{fig:J_scheme}
\end{figure}

At this point let us remember that the $J$ values depend mostly on (i) the
distance between the two magnetic atoms (M), (ii) the distance between the
magnetic atoms and the bridging ligands (L) and (iii) the
$\widehat{\text{M-L-M}}$ angle between the magnetic atoms and the ligands.
The Fe$^{2+}$ ($3d^6$ configuration) are the magnetic atoms, and their
magnetic interactions are bridged by two S ligands (see
Fig.~\ref{fig:J_scheme}).
The three modes exhibiting anomalies at $T_N$ mostly involve movements of the
sulfur atoms bridging the interactions along the ladder legs (see SI
Fig.~\ref{SM_fig:Anomaly_Ts_DFT_144} and~\ref{SM_fig:Anomaly_Tn_315_Eb}).  As
a consequence, these modes are expected to be coupled to the magnetic exchange
along the ladder-legs.  The same analysis holds for the \unit[124]{cm$^{-1}$}
mode seen when the light polarization is along $\vec b$ (see SI
Fig.~\ref{SM_fig:Anomaly_Ts_DFT_124}b).  The \unit[124]{cm$^{-1}$} mode seen
when the light polarization is along $\vec a$ is different as it mostly
involves the movement of the Fe ions (see Fig~\ref{fig:J_scheme}).  On the
ladder-rungs it results in a large change in the distances between the Fe and
the bridging-S as well as the $\widehat{\text{Fe-S-Fe}}$ angles. On the
ladder-legs the main effects are on the Fe-Fe distance and the distances
between the Fe and the bridging-S. In both cases we expect a strong coupling
between this phonon mode and both the intra-ladder exchange integrals.

These intra-ladder interactions are expected to be much larger than the
inter-ladder ones. As the latters are responsible for the magnetic transition
at $T_N$, one can expect the existence of intra-ladder short-range
magnetic correlations above $T_N$. In view of those two observations one can
wonder whether the structural transition at $T_S$ could originate from the
intra-ladder short-range magnetic correlations.

In order to confirm this hypothesis we performed neutron scattering to study
the behavior of magnetic Bragg's peaks through the two phase transitions.

\subsection{Magnetic Fluctuations}
To investigate the magnetic fluctuations and extract the associated
correlation lengths, we performed momentum ($\mathbf{Q}$) and energy
($\hbar\omega$) scans around the magnetic Bragg position
$\mathbf{Q} = (\frac{1}{2}, \frac{1}{2}, 1)$. Momentum scans were carried out
both along the ladder direction, i.e. $(\nicefrac{1}{2}, \nicefrac{1}{2}, L)$,
and perpendicular to it by simultaneously varying the $H$ and $K$ components
along $(H,H,1)$. In the following, these measurements are referred to as $L$,
$HH$, and $E$ scans, respectively. All measurements were performed under
identical experimental conditions over the temperature range
$10~\mathrm{K} \leq T \leq 140~\mathrm{K}$.

\begin{figure}[htbp]
    \centering
    \includegraphics[width=0.5\textwidth]{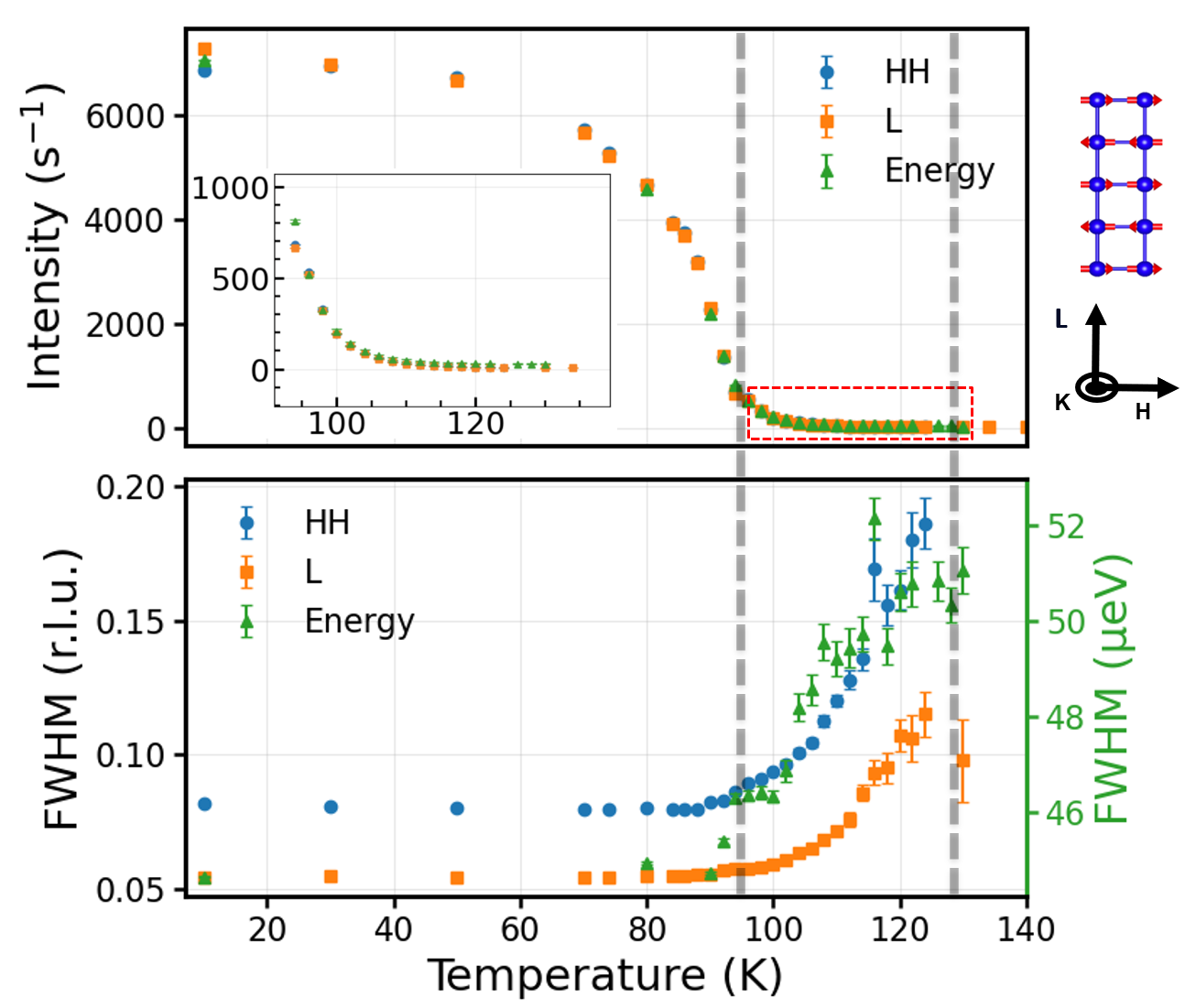}
    \caption{ \textbf{Neutron scattering.}  maximum intensity, $A$ of eq.~2 in
      SI~\cite{SI} (top) and FWHM (bottom) evolutions as a function of temperature for
      $HH$, $L$ and $E$ scans around the magnetic Bragg peak
      $(\nicefrac{1}{2},\nicefrac{1}{2},1)$.  Vertical gray dashed lines mark
      $T_{N}$ and $T_{S}$, respectively. Inset in top panel displays a zoom in
      for $T_N\le T\le T_S$.}
    \label{fig:Mag}
\end{figure}

The peaks were fitted using pseudo-Voigt functions along each direction (see
Supplemental Information for details). At the lowest temperature
(\unit[T=10]{K}), the energy linewidth has a full
width at half maximum (FWHM) of \unit[44]{$\mu$eV}. The maximum intensities extracted from the three
types of scans are plotted as a function of temperature in the top panel of
Fig.~\ref{fig:Mag}. Within experimental uncertainty, the intensity and its
temperature dependence are independent of the scan direction.

The temperature evolution of the FWHM along the ladder direction,
perpendicular to the ladder, and in energy is shown in the bottom panel of
Fig.~\ref{fig:Mag}. At low temperatures and up to $T_N \approx 95$~K, the
widths remain constant. The $\mathbf{Q}$-widths are limited by the crystal
mosaicity, with a slightly narrower width along the ladder direction,
consistent with the quasi-one-dimensional character of the system and previous
reports on sibling compound \bfse~\cite{Roll2023}. The energy width remains
also constant at $44~\mu\mathrm{eV}$.

Above $T_N$, a progressive broadening is observed both in momentum and
energy. The increase of the $\mathbf{Q}$-width indicates a reduction of the
magnetic correlation lengths along and perpendicular to the ladder direction,
while the broadening in energy reflects a shortening of the magnetic
fluctuation lifetime.  We note that a quantitative extraction of these
  correlation lengths and lifetimes from triple-axis data would require a full
  characterization of the four-dimensional resolution ellipsoid across
  multiple sample orientations, a measurement program well beyond the scope of
  the present study. However, the qualitative observation of finite
  broadenings clearly exceeding the instrumental resolution is sufficient to
  establish our conclusions.  The magnetic signal becomes undetectable above
$T_S\approx \unit[125-130]{K}$, suggesting that the magnetic correlations are
no longer spatially coherent in reciprocal space and that the associated
spectral weight is distributed over a broad momentum range and thus is
assimilated as a homogenous background component.

Overall, these inelastic neutron scattering results demonstrate that the
magnetic order is three-dimensional, long-ranged, and static below
$T_N$. Between $T_N$ and $T_S$, the system exhibits short-range,
three-dimensional dynamic correlations (and not one-dimensional as expected),
and above $T_S$ no measurable three-dimensional magnetic correlations are
detected.

\section{Discussion and interpretation}

\begin{figure}[htbp]
  \centering
  \includegraphics[width=1.\linewidth]{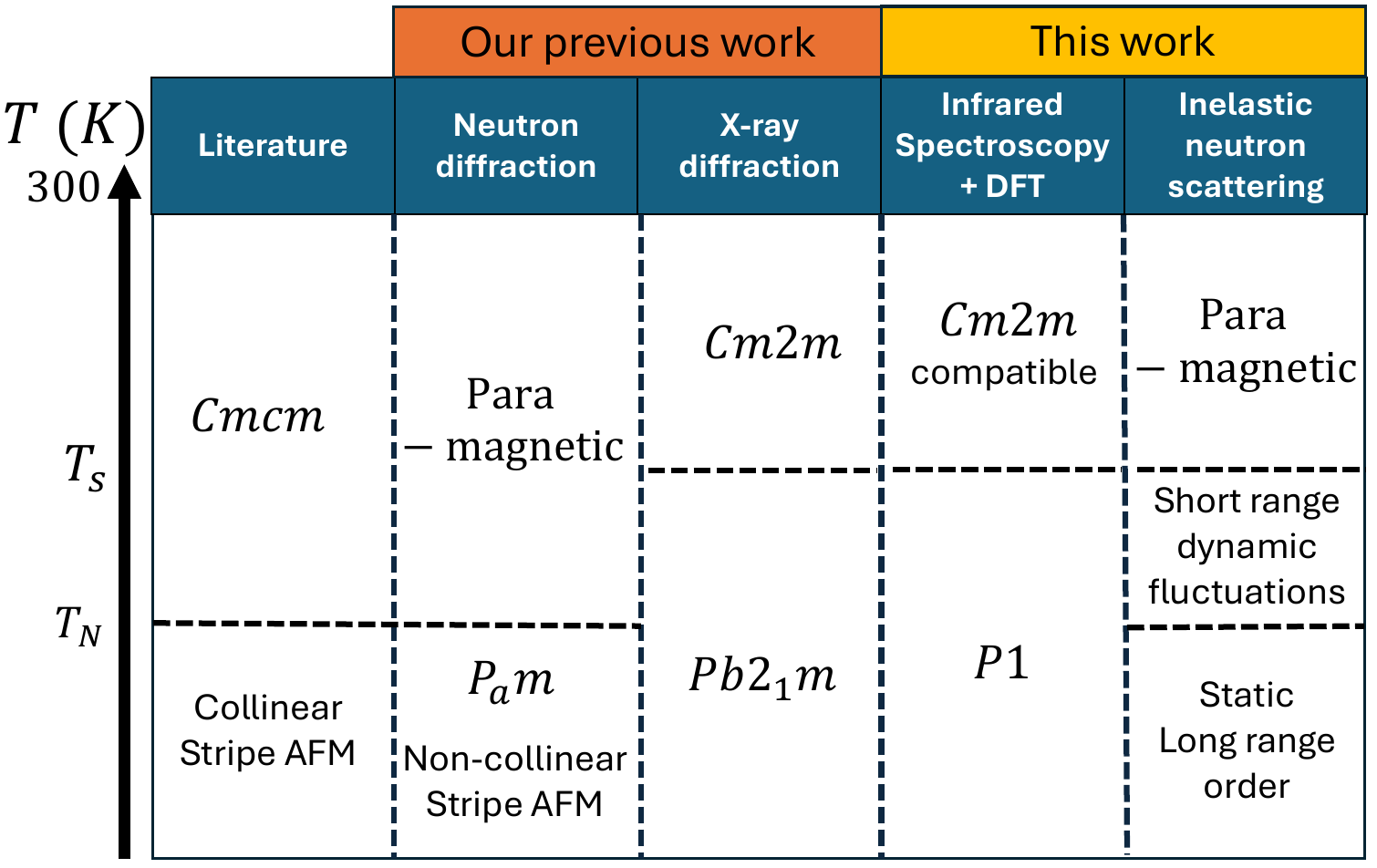}
  \caption{Summary of the multi-technique ambient pressure
    temperature-dependent results. The diagram compares the structural and
    magnetic information extracted from x-ray diffraction, neutron diffraction
    (our previous work~\cite{Oubaid2026BaFe2S3}), infrared spectroscopy,
    lattice dynamics and inelastic neutron scattering (this work) and
    contrasts them with the previous literature consensus
    \cite{Takahashi2015,Zheng2018,IR_bafe2sse3_prb}.}
    \label{fig:Conclusion}
\end{figure}

Our work led to several unexpected results. First, the symmetry analysis of
the infrared-active phonons, supported by first-principles calculations,
demonstrates that the crystal symmetry is lower than the previously proposed
$Pb2_1m$ space group and is only consistent with a $P1$ description at low
temperature. The loss of residual point-group symmetry operations is required
to account for the full set of experimentally observed phonon modes and their
polarization dependence. Second, neutron scattering establishes that below
$T_N \approx$ \unit[95]{K} the magnetic order is three-dimensional,
long-ranged, and static.  Between $T_N$ and $T_S \approx \unit[125-130]{K}$, the
system exhibits three-dimensional, short-range dynamic magnetic correlations,
as evidenced by the broadening in momentum and energy both along and
perpendicular to the ladder direction. Above $T_S$, no measurable magnetic
correlations persist, indicating that the magnetic spectral weight has been
diluted. Whether this dilution is isotropic in all three momentum directions
and in energy~--- as expected for a purely paramagnetic state~--- or whether
one-dimensional fluctuations persist above $T_S$ but remain undetected cannot
be determined from the present data.  As a previous study on the neighboring
compound~\cite{Roll2023} suggests dominant intra-ladder exchange interactions,
the latter scenario appears more likely; however, any residual spectral weight
confined along the ladder direction would be spread over the $(H,H,0)$ plane
and thus diluted below the detection threshold given the available statistical
precision.  Third, the temperature evolution of selected phonon modes reveals
anomalies at both $T_N$ and $T_S$. The phonon anomalies at $T_N$ reflect
magnetoelastic coupling associated with the onset of static long-range
magnetic order. The modes affected at $T_S$ predominantly involve either Fe or
S atomic displacements. These motions modulate the Fe--Fe exchange paths along
the rungs and legs of the ladders and should have a direct impact on the
magnetic properties.

Remarkably, the phonon anomalies observed at $T_S$ occur at the same
temperature as the onset of three-dimensional short-range dynamical magnetic
correlations detected by neutron scattering.  These phonon anomalies thus
establish a bridge between the static structural transition~--- at which
lattice anomalies are naturally expected~--- and the appearance of magnetic
fluctuations, since the atomic displacements associated with these modes
directly modulate the magnetic exchange pathways.

This phenomenology is reminiscent of the iron-pnictide 122 family, in
particular BaFe$_2$As$_2$, where a tetragonal-to-orthorhombic structural
transition occurs at or slightly above the stripe-type antiferromagnetic
ordering temperature \cite{Kim2011}. In that case, extensive experimental
\cite{Kimber2009,Huang2008,Wu2014} and theoretical work
\cite{fang2008theory,Fernandes2010effects,Fernandes2014} has established that
the structural distortion is electronically driven and closely tied to
spin-driven nematic fluctuations.

The situation in \bfs\ shares key similarities with this phenomenology, though
important differences must be noted.  As in the 122 pnictides, the structural
transition precedes the establishment of static long-range magnetic order and
coincides with the onset of magnetic correlations. However, unlike
BaFe$_2$As$_2$, where the instability is often described in terms of
electronic nematic order in an itinerant metallic system, \bfs\ is described
as a Mott insulator with no electronic nematicity (there is no four-fold
symmetry to break in this case). In \bfs\, the structural transition occurs in
a regime where magnetic correlations become three-dimensional but remain
dynamic and short-ranged. This suggests that magnetoelastic coupling
associated with correlated local moments could be the driving mechanism of the
symmetry lowering at $T_S$.

\section{Conclusion}
In this work, we combined polarized synchrotron infrared spectroscopy,
hybrid-functional DFT lattice-dynamics calculations, and inelastic neutron
scattering to elucidate the interplay between structure, lattice dynamics, and
magnetism to reveal the ground state of \bfs.  This highlights the subtle
coupling between magnetic short-range order and lattice degrees of freedom as
well as its central role in shaping the phase diagrams of iron-based
superconductors.

\section{Acknowledgment}
We acknowledge the MORPHEUS platform at the Laboratoire de Physique des
Solides for sample orientation and alignment. We acknowledge SOLEIL for
providing the synchrotron beamtime (proposals : 20230487,20242024) and ILL the
neutron beamtime (DOI: 10.5291/ILL-DATA.4-03-1780).  M.-B.L. and
S.D. aknowledge the IDRIS high-performance computer center for the calculation
time under the GENCI project n$^\circ$0801842. This work was financially
supported by the ANR COCOM 20-CE30-0029, the France 2030 program
ANR-11-IDEX-0003 via Integrative Institute of Materials from Paris-Saclay
University - 2IM@UPSaclay, the Paris Ile-de-France Region in the framework of
DIM MaTerRE (project DAC-VX).

\bibliography{ref}

\end{document}


\pagenumbering{arabic}

\title{Supplemental Material (SM) for:\\
``Ground State of {\bfs} from Lattice and Spin Dynamics''}

\author{Y. Oubaid}
\thanks{These authors contributed equally to this work.}
\affiliation{Universit\'e Paris-Saclay, CNRS, Laboratoire de Physique des Solides, 91405, Orsay, France.}
\affiliation{Synchrotron SOLEIL, L\'Orme des Merisiers, Saint Aubin BP 48, 91192, Gif-sur-Yvette, France}

\author{S. Deng}
\thanks{These authors contributed equally to this work.}
\affiliation{Institut Laue-Langevin, 71 avenue des Martyrs, 38042 Grenoble, France}

\author{N.S. Dhami}
\affiliation{Universit\'e Paris-Saclay, CNRS, Laboratoire de Physique des Solides, 91405, Orsay, France.}

\author{M. Verseils}
\affiliation{Synchrotron SOLEIL, L\'Orme des Merisiers, Saint Aubin BP 48, 91192, Gif-sur-Yvette, France}
\author{P. Fertey}
\affiliation{Synchrotron SOLEIL, L\'Orme des Merisiers, Saint Aubin BP 48, 91192, Gif-sur-Yvette, France}
\author{P. Steffens}
\affiliation{Institut Laue-Langevin, 71 avenue des Martyrs, 38042 Grenoble, France}

\author{D. Bounoua}
\affiliation{Universit\'e Paris-Saclay, CNRS-CEA, Laboratoire L\'eon Brillouin, 91191, Gif sur Yvette, France}

\author{A. Forget}
\affiliation{Universit\'e Paris-Saclay, CEA, CNRS, SPEC, 91191, Gif-sur-Yvette, France.}

\author{D. Colson}
\affiliation{Universit\'e Paris-Saclay, CEA, CNRS, SPEC, 91191, Gif-sur-Yvette, France.}

\author{P. Foury-Leylekian} \affiliation{Universit\'e Paris-Saclay, CNRS,
  Laboratoire de Physique des Solides, 91405, Orsay, France.}

\author{M.B. Lepetit}
\email[Corresponding author:]{Marie-Bernadette.Lepetit@neel.cnrs.fr}
\affiliation{Institut N\'eel, CNRS, 38042 Grenoble, France}
\affiliation{Institut Laue-Langevin, 71 avenue des Martyrs, 38042 Grenoble, France}

\author{V. Bal\'edent}
\email[Corresponding author:]{victor.baledent@universite-paris-saclay.fr}
\affiliation{Universit\'e Paris-Saclay, CNRS, Laboratoire de Physique des Solides, 91405, Orsay, France.}
\affiliation{Institut universitaire de France (IUF)}

\date{\today}

\maketitle
\tableofcontents

\section{Infrared Spectroscopies and Phonon Modes Fitting}

Experimental phonon frequencies were obtained by fitting the data using the usual Drude–Lorentz model for the dielectric function of insulating materials \cite{kuzmenko2005kramers}. The dielectric function is thus expressed as the sum of harmonic oscillators, as described by the equation \ref{Drude_lorentz}  :

\begin{equation}
\label{Drude_lorentz}
    \epsilon(\omega) = \epsilon_{\infty} + \sum_{k} \frac{\omega_{p,k}^2 }{\omega_{0,k}^2 - \omega^2 - i\Gamma_k\, \omega}
\end{equation}

where $\epsilon_{\infty}$ is the dielectric constant at infinite frequency,
and $\omega_{0,k}$, $\omega_{p,k}$, and $\Gamma_{k}$ represent respectively
the  oscillator frequency, plasma frequency and energy width of the $k^{th}$
harmonic oscillator.  The model parametrizes the complex permittivity as shown
in the equation \ref{Drude_lorentz}.  To asses the quality of the fit we
defined the goodness of fit (GOF) parameter quantified by :
  \[
    \mathrm{GOF} = 1 - \frac{\mathrm{SS}_{\mathrm{res}}}{\mathrm{SS}_{\mathrm{tot}}},
    \qquad
    \mathrm{SS}_{\mathrm{res}}=\sum_{i=1}^{n}\!\bigl(y_i - \epsilon(\omega_i) \bigr)^2,\quad 
  \]
  \[
    \mathrm{SS}_{\mathrm{tot}}=\sum_{i=1}^{n}\!\bigl(y_i-\bar y\bigr)^2 \qquad \text{with} \qquad 
    \bar y= \sum_{i=1}^{n}\frac{y_i}{n}
  \]
  where, $n$ is the number of measured points, $y_i$ is the measured
  reflectance at $\omega_i$,  $\epsilon(\omega_i)$ is the Drude--Lorentz fit
  at $\omega_i$ (eq.~\ref{Drude_lorentz}).
  
    The average value for the fits was
  around $\mathrm{GOF}\approx 0.99$.

\begin{figure}[htbp]
    \centering
    \includegraphics[width=0.5\textwidth]{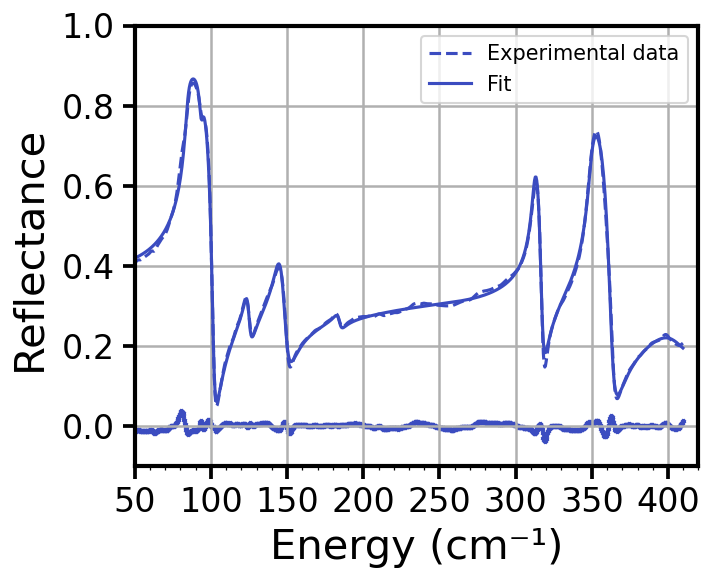}
    \caption{Reflectance fitting using a Drude--Lorentz (DL) dielectric model.
Measured reflectance in the rung direction of {\bfs}
($\mathbf{E}\!\parallel\! a$, 30~K; solid line) together with the DL fit (dashed).
The residuals (data $-$ fit) are shown in the same panel as a curve centered around zero.
}
    \label{SM_fig:fitting_proof}
\end{figure}

\clearpage
\section{Lattice Dynamics Calculations}

\begin{table}[htbp]
  \centering
  \caption{DFT frequencies (cm$^{-1}$) of optical phonons computed in $Pm$, $Pc$, and $P2_1$ space groups, and their
     best experimental assignment.}
  \label{tab_SM:Pm_Pc_P21_Ph}
\noindent
\begin{minipage}{0.25\linewidth}
  \centering $Pm$
\end{minipage}
\hspace{0.75cm}
\begin{minipage}{0.25\linewidth}
  \centering $Pc$
\end{minipage}
\hspace{0.75cm}
\begin{minipage}{0.25\linewidth}
  \centering $P2_1$
\end{minipage}\\
  \begin{minipage}[t]{0.15\linewidth}
    \vspace{0pt}
    \begin{tabular}{ccc}
      \hline \hline
      DFT   & \multicolumn{2}{c}{IR} \\
      $A'$ & {\small $\vec E\!\parallel\! \vec a$} & {\small $\vec E\!\parallel\! \vec b$}\\
      \hline
      31.5  &      &     \\
      40.7  &      &     \\
      48.8  &      &     \\
      53.6  &      &     \\
      55.7  &      &     \\
      73.0  &      &     \\
      75.0  &      &     \\
      80.2  &      &     \\
      83.4  & 84   &     \\
      85.3  &      &     \\
      86.1  &      &     \\
      92.2  & 93   &     \\
      93.9  &      & 95  \\
      113.8 &      &     \\
      118.4 & 124  & 124 \\
      131.2 &      &     \\
      135.0 &      &     \\      
      141.8 & 144  &     \\
      150.0 &      & 145 \\
      155.3 &      &     \\
      166.9 &      &     \\
      167.4 &      &     \\
      196.7 & 183  &     \\
      198.0 &      &     \\
      202.1 &      &     \\
      211.7 &      &     \\
      255.8 &      &     \\
      257.9 &      &     \\
      271.0 &      &     \\
      276.4 &      &     \\
      278.0 &      &     \\
      278.5 &      &     \\
      280.9 &      &     \\
      292.3 &      &     \\
      311.1 & 312  & 312 \\
      329.5 & 329  & 329 \\
      338.8 &      &     \\
      340.0 & 349  & 355 \\
      374.1 &      &     \\
      377.5 &      &     \\
      379.0 &      &     \\
      393.9 & 404  & 404 \\
      \hline \hline
     \end{tabular}
   \end{minipage}
   \begin{minipage}[t]{0.1\linewidth}
     \vspace{0pt}
   \begin{tabular}{cc}
      \hline \hline
      DFT   & IR \\
      $A''$ & {\small $\vec E\!\parallel\! \vec c$}\\
      \hline
     27.7  &   \\
     40.9  &   \\
     43.1  &   \\
     44.3  &   \\
     49.5  &   \\
     88.5  & 79\\
     94.2  &   \\
     97.5  &   \\
     104.3 &   \\
     105.2 &   \\
     107.9 &   \\
     109.2 &   \\
     110.3 & 117 \\
     133.2 &     \\
     134.3 &     \\
     212.6 &     \\
     224.0 &     \\
     225.7 &     \\
     229.7 &     \\
     241.6 &     \\
     246.2 &   \\
     248.2 &   \\
     252.3 &   \\
     274.3 &   \\
     275.4 &   \\
     276.2 &   \\
     277.5 &   \\
           & \tred{\bf 329}\\
           & \tred{\bf 404}\\
      \hline \hline
     \end{tabular}
   \end{minipage} 
   \hspace{0.75cm}
  \begin{minipage}[t]{0.15\linewidth}
    \vspace{0pt}
    \begin{tabular}{ccc}
      \hline \hline
      DFT   & \multicolumn{2}{c}{IR} \\
      $A'$ & {\small $\vec E\!\parallel\! \vec a$} & {\small $\vec E\!\parallel\! \vec b$}\\
      \hline
      28.1  &      &     \\
      41.2  &      &     \\
      44.7  &      &     \\
      49.1  &      &     \\
      51.4  &      &     \\
      59.6  &      &     \\
      82.2  &      &     \\
      84.9  & 84   &     \\
      93.7  & 93   &     \\
      95.2  &      & 95  \\
      99.9  &      &     \\
      108.9 &      &     \\
      109.2 &      &     \\
      130.4 & 124  & 124 \\
      134.5 &      &     \\
      134.7 &      &     \\      
      142.0 & 144  & 145 \\
      166.0 &      &     \\
      196.9 & 183  &     \\
      211.7 &      &     \\
      217.5 &      &     \\
      227.2 &      &     \\
      246.1 &      &     \\
      247.8 &      &     \\
      258.4 &      &     \\
      269.6 &      &     \\
      275.8 &      &     \\
      276.8 &      &     \\
      277.6 &      &     \\
      280.9 &      &     \\                    
      310.5 & 312  & 312 \\
      \multirow{2}*{338.3} 
            & \tred{\bf 329} & \tred{\bf 329}\\
            & \tred{\bf 349} & \tred{\bf 355}\\
      378.4 &      &     \\
      378.9 &      &     \\
            & \tred{\bf 404} & \tred{\bf 404} \\
      \hline \hline
     \end{tabular}
   \end{minipage}
   \begin{minipage}[t]{0.1\linewidth}
     \vspace{0pt}
   \begin{tabular}{cc}
      \hline \hline
      DFT   & IR \\
      $A''$ & {\small $\vec E\!\parallel\! \vec c$}\\
      \hline
    30.4  &   \\
    39.9  &   \\
    43.1  &   \\
    50.1  &   \\
    54.9  &   \\
    73.5  &   \\
    74.9  &   \\
    80.7  & 79\\
    86.8  &   \\
    92.3  &   \\
    97.2  &   \\
    98.4  &   \\
    104.9 &   \\
    108.3 &   \\
    118.1 & 117\\
    135.0 &   \\
    148.9 &   \\
    154.8 &   \\
    167.4 &   \\
    197.5 &   \\
    201.6 &   \\
    224.6 &   \\
    227.4 &   \\
    241.7 &   \\
    253.3 &   \\
    256.5 &   \\
    274.3 &   \\
    278.0 &   \\
    278.4 &   \\
    279.0 &   \\
    292.0 &   \\
    329.6 & 329 \\
    339.3 &   \\
    374.5 &   \\
    394.4 & 404 \\
      \hline \hline
     \end{tabular}
   \end{minipage} 
   \hspace{0.75cm}
   \begin{minipage}[t]{0.15\linewidth}
     \vspace{0pt}
    \begin{tabular}{ccc}
      \hline \hline
      \multicolumn{1}{c}{DFT} & \multicolumn{2}{c}{IR} \\
      $B$ & {\small $\vec E\!\parallel\! \vec a$} & {\small $\vec E\!\parallel\! \vec b$}\\
      \hline
      28.6  &      &     \\
      40.4  &      &     \\
      43.7  &      &     \\
      44.6  &      &     \\
      76.8  &      &     \\
      81.4  &      &     \\
      82.7  &      &     \\
      84.4  & 84   &     \\
      93.2  & 93   &     \\
      96.6  &      & 95  \\
      103.1 &      &     \\
      105.9 &      &     \\
      108.8 &      &     \\
      131.4 & 124  & 124 \\
      133.7 &      &     \\
      142.6 & 144  & 145 \\
      150.0 &      &     \\
      155.4 &      &     \\
      197.6 & 183  &     \\
      212.1 &      &     \\
      224.4 &      &     \\
      225.7 &      &     \\
      246.1 &      &     \\
      253.1 &      &     \\
      255.6 &      &     \\
      269.4 &      &     \\
      275.2 &      &     \\
      277.3 &      &     \\
      277.9 &      &     \\
      280.3 &      &     \\
      310.7 & 312  & 312 \\
      \multirow{2}*{339.5} 
            & \tred{\bf 329} & \tred{\bf 329}\\
            & \tred{\bf 349} & \tred{\bf 355}\\
      378.8 &      &     \\
      394.3 & 404  & 404 \\
      \hline \hline
     \end{tabular}
   \end{minipage}
   \begin{minipage}[t]{0.1\linewidth}
     \vspace{0pt}
   \begin{tabular}{cc}
      \hline \hline
      \multicolumn{1}{c}{DFT}   & \multicolumn{1}{c}{IR} \\
    $A$ & {\small $\vec E\!\parallel\! \vec c$}\\
      \hline
     29.8  &   \\
     40.4  &   \\
     48.9  &   \\
     49.5  &   \\
     52.4  &   \\
     54.1  &   \\
     74.8  &   \\
     80.8  & 79 \\
     85.9  &   \\
     93.4  &   \\
     94.2  &   \\
     101.6 &   \\
     108.2 &   \\
     110.2 &   \\
     119.4 & 117 \\
     133.1 &   \\
     135.2 &   \\
     167.0 &   \\
     167.9 &   \\
     197.1 &   \\
     202.0 &   \\
     223.5 &   \\
     225.5 &   \\
     241.4 &   \\
     250.2 &   \\
     257.6 &   \\
     273.8 &   \\
     275.7 &   \\
     276.4 &   \\
     277.4 &   \\
     291.8 &   \\
     329.4 & 329 \\
     338.2 &   \\
     374.4 &     \\
     378.4 &     \\
           & \tred{\bf 404}\\
      \hline \hline
   \end{tabular}
 \end{minipage}
\end{table}

\clearpage
\section{Phonon Temperature-dependence Anomalies and displacements}

\begin{figure}[htbp]
    \centering
    {(a) DFT modes corresponding to IR mode $\vec E\!\parallel\! \vec a$}\\ 
    \includegraphics[width=0.3\textwidth]{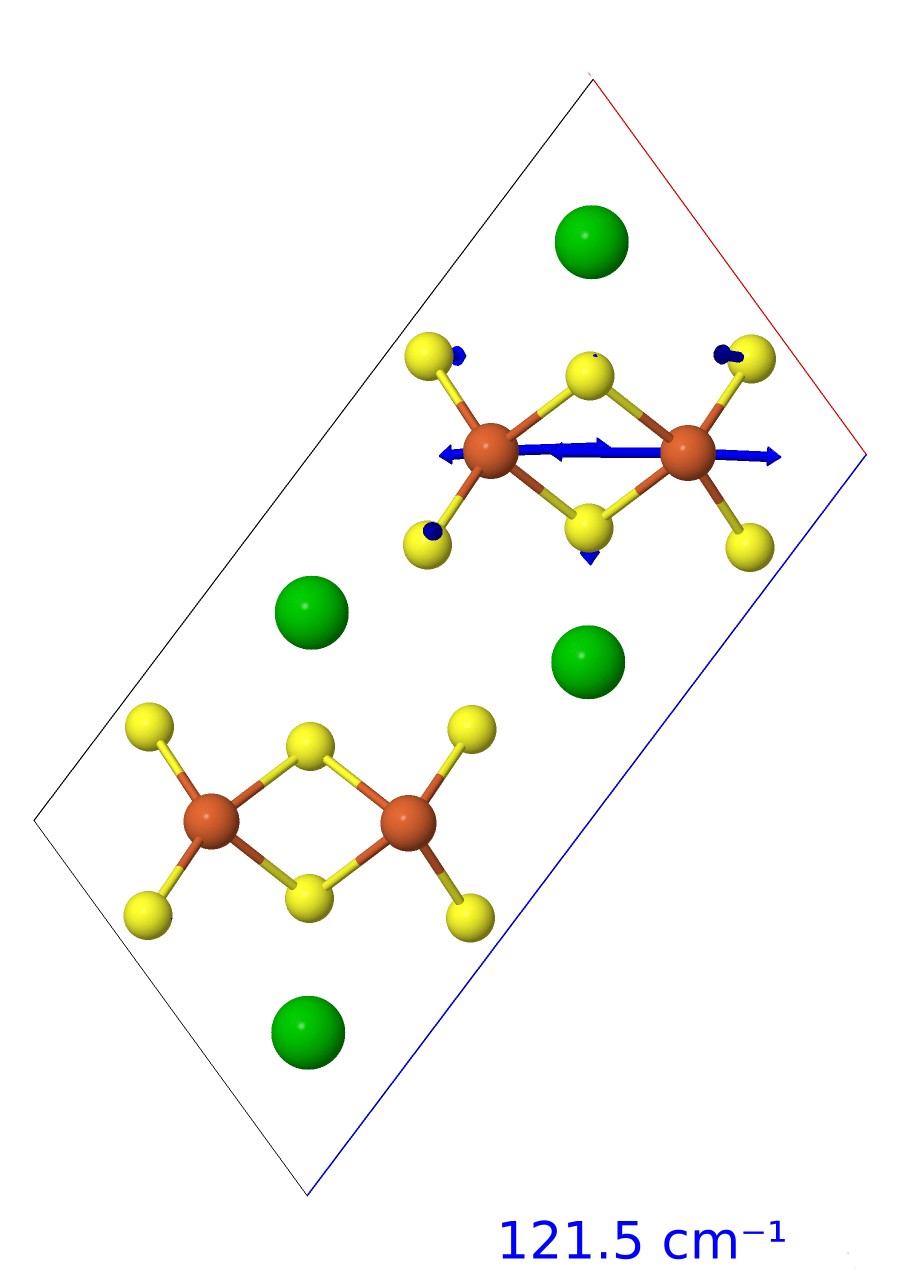}
    \includegraphics[width=0.3\textwidth]{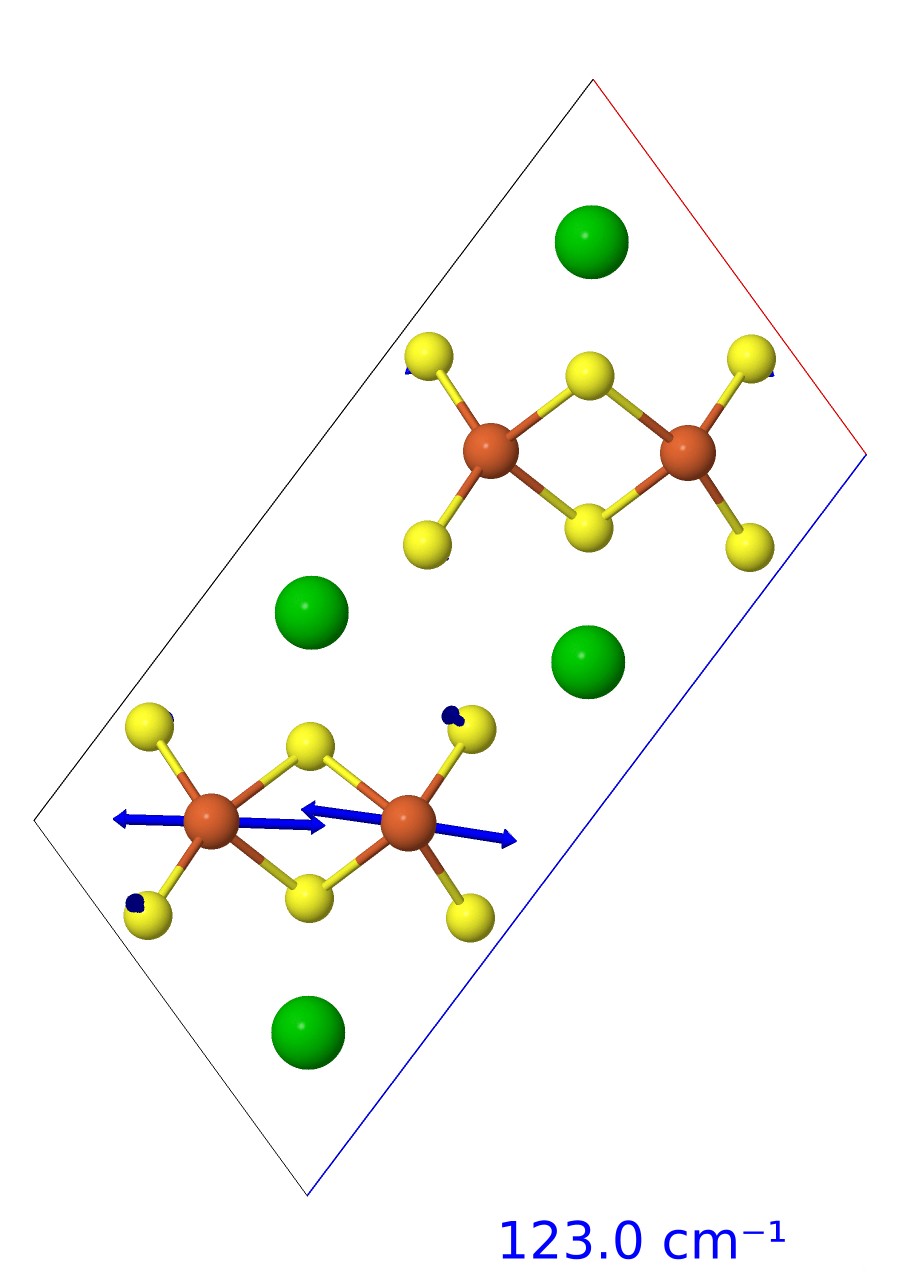}
    \includegraphics[width=0.2\textwidth]{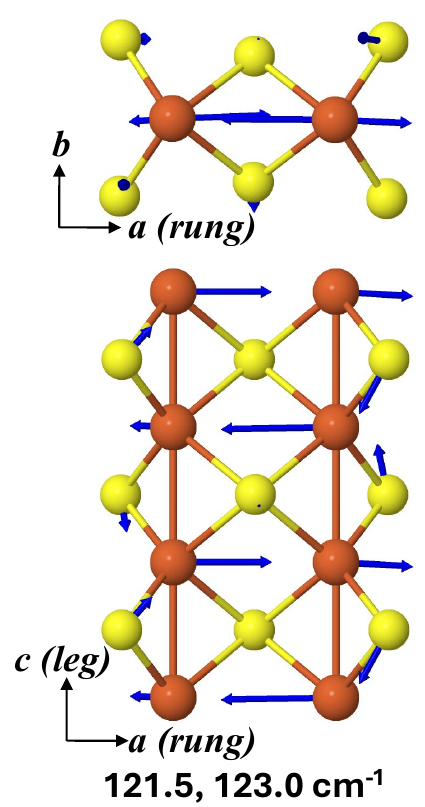} \\
    \vspace{2.0em}
    {(b) DFT modes corresponding to IR mode $\vec E\!\parallel\! \vec b$}\\
    \includegraphics[width=0.3\textwidth]{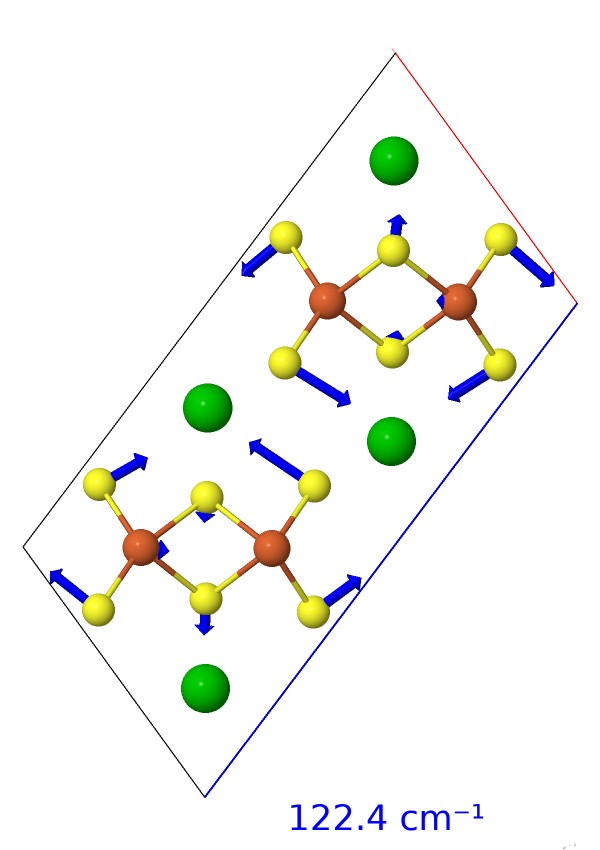}
    \includegraphics[width=0.3\textwidth]{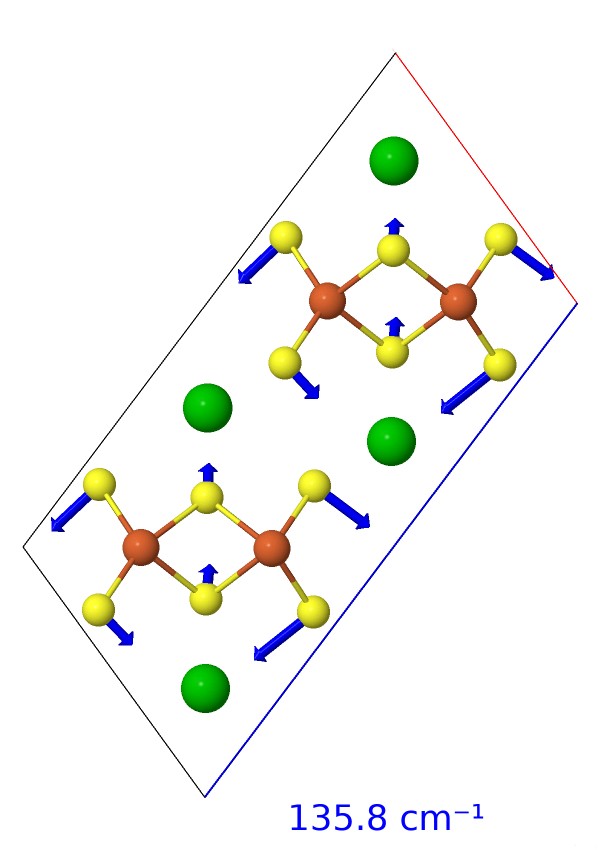}
    \includegraphics[width=0.18\textwidth]{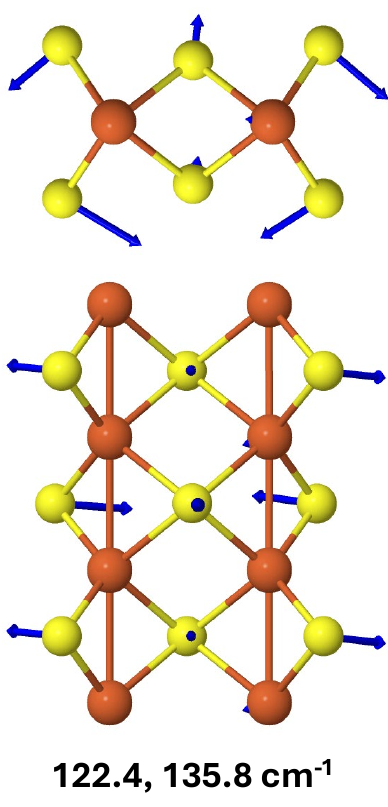}
    \caption{
    Atomic displacement vectors for DFT-phonon modes associated with the IR mode seen at \unit[124]{cm$^{-1}$}, as well as their counterpart issued from the symmetry breaking, when (a) $\vec E\!\parallel\! \vec a$ and (b) $\vec E\!\parallel\! \vec b$. 
    There are two degenerate modes for each direction, corresponding to the fact that there are two independent Fe-ladders in the crystalline structure. 
    On the right panels, we show the displacement of one ladder as representative, in reference to the light polarization direction in the $Cmcm$ notation. 
    (a) Both modes can be seen in IR, and the main displacement is clearly along the ladder-rung direction, thus corresponding to the IR observables when $\vec E\!\parallel\! \vec a$. 
    (b) One mode is symmetric while the other is antisymmetric. We expect to see the antisymmetric one in IR. The net displacements along ladder-rung direction are negligible and the main displacements are along the perpendicular direction to the ladders. We hence assign this mode to the IR observables when $\vec E\!\parallel\! \vec b$.}
  \label{SM_fig:Anomaly_Ts_DFT_124}
\end{figure}

\begin{figure}[h]
  \centering
  \includegraphics[width=0.45\textwidth]{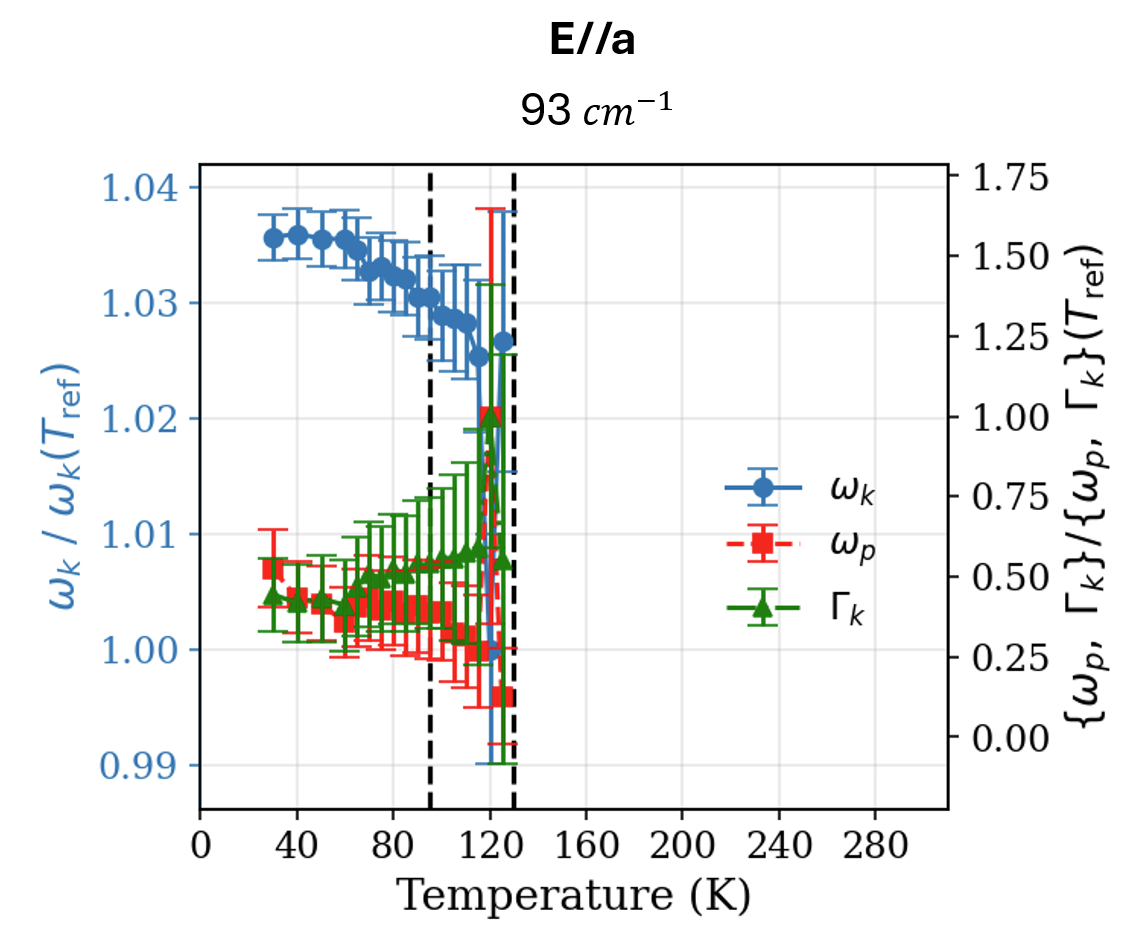}
  \caption{Temperature dependance of the phonon seen at
    \unit[93]{cm$^{-1}$}. Blue~: oscillator frequencies, $\omega_k$.  Red~:
    normalized plasma frequencies, $\omega_p$.  Green~: width,
    $\Gamma_k$.}
    \label{SM_fig:93cm}
  \end{figure}

\begin{figure}[h]
    \centering
    \includegraphics[width=0.38\textwidth]{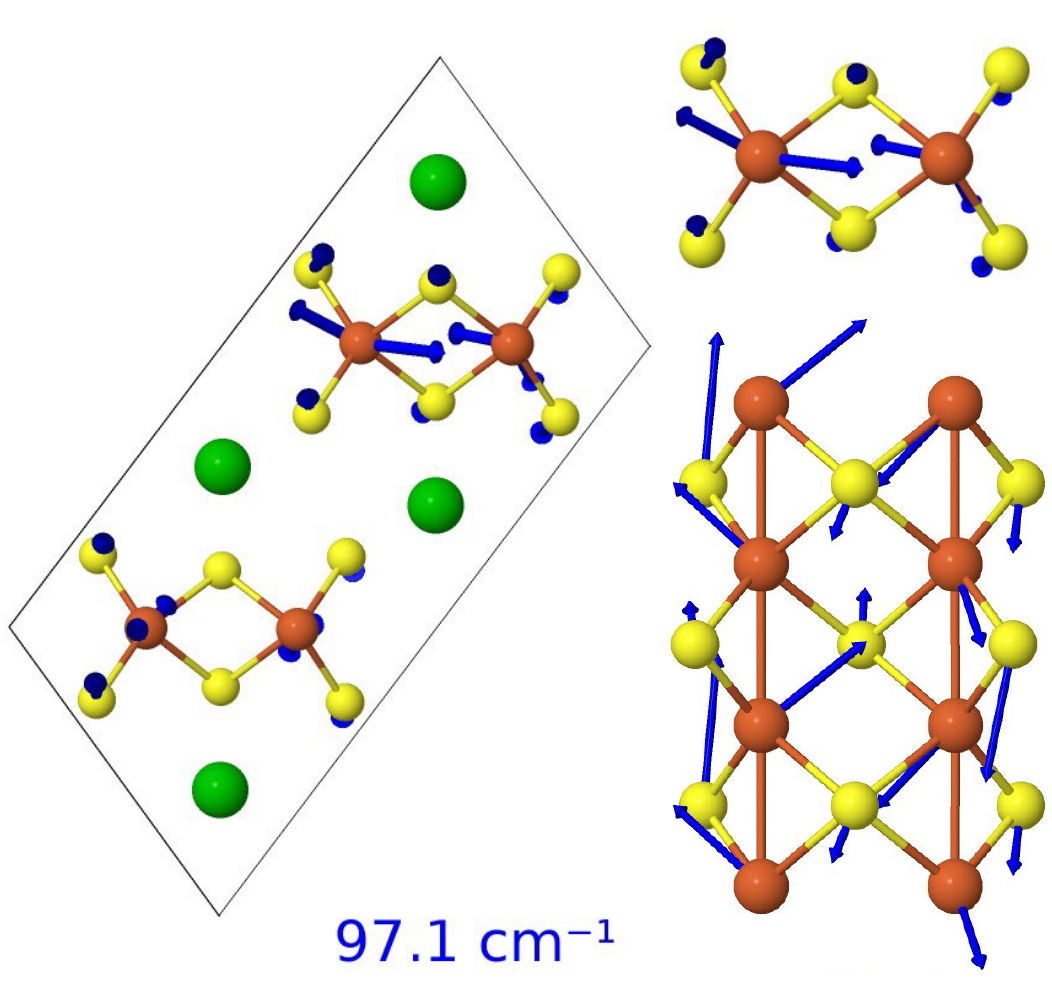} \hspace{12pt}
    \includegraphics[width=0.25\textwidth]{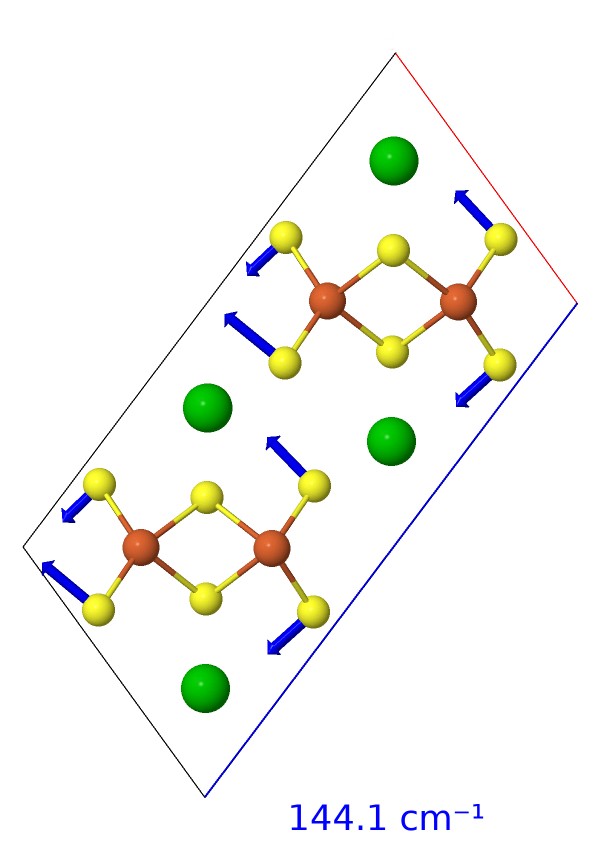}
    \includegraphics[width=0.22\textwidth]{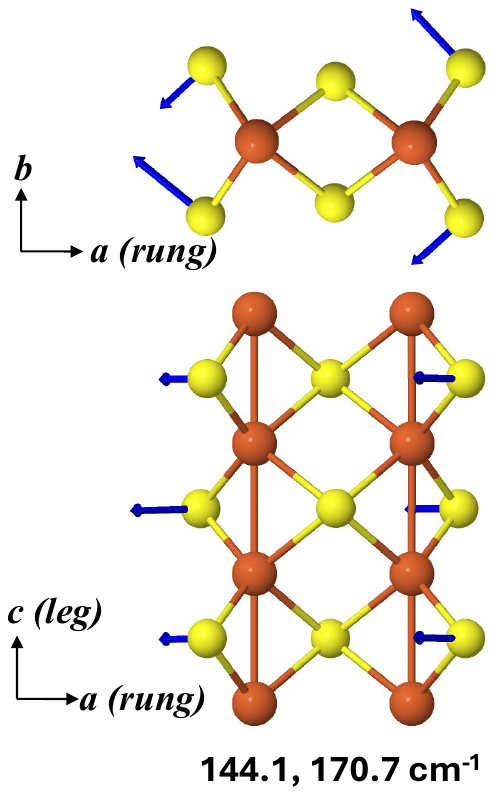} 
    \caption{Atomic displacement vectors for DFT-phonon mode, assigned with
      the \unit[93]{cm$^{-1}$} IR mode (seen when $\vec E \parallel \vec a$), and the \unit[144]{cm$^{-1}$}/\unit[145]{cm$^{-1}$} IR mode seen when the light polarisation is either along
      $\vec a$ or $\vec b$.}
    \label{SM_fig:Anomaly_Ts_DFT_144}
\end{figure}

\begin{figure}[h]
  \centering
    \includegraphics[width=0.18\textwidth]{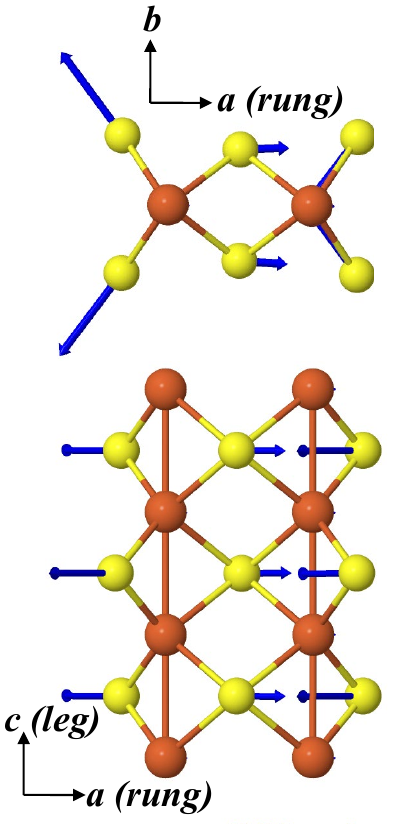} \hspace{12pt}
    \includegraphics[width=0.25\textwidth]{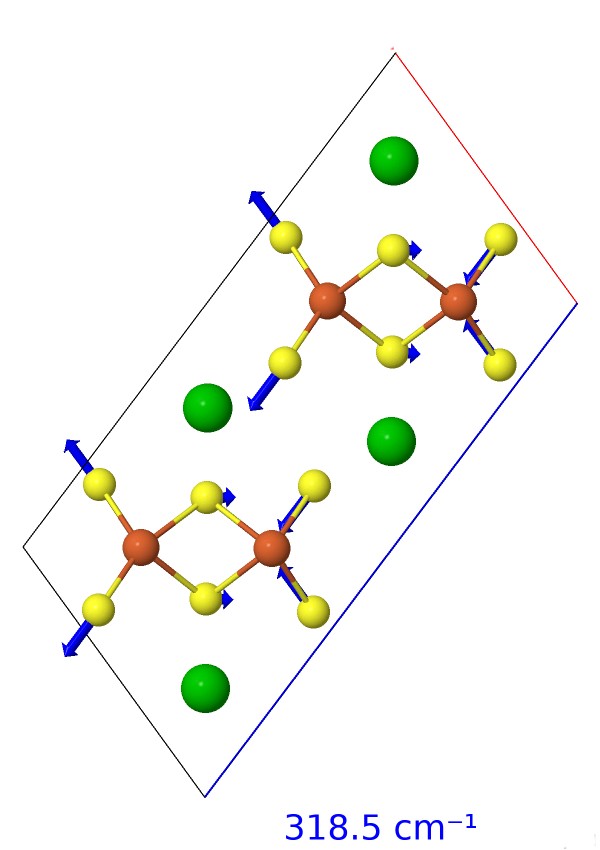}
    \caption{Atomic displacement vectors for DFT-phonon modes associated with
      the IR \unit[315]{cm$^{-1}$} mode.}
    \label{SM_fig:Anomaly_Tn_315_Eb}
\end{figure}

\clearpage
\section{Inelastic Neutron scattering}
\begin{figure}[htbp] 
    \centering
    \includegraphics[width=1\textwidth]{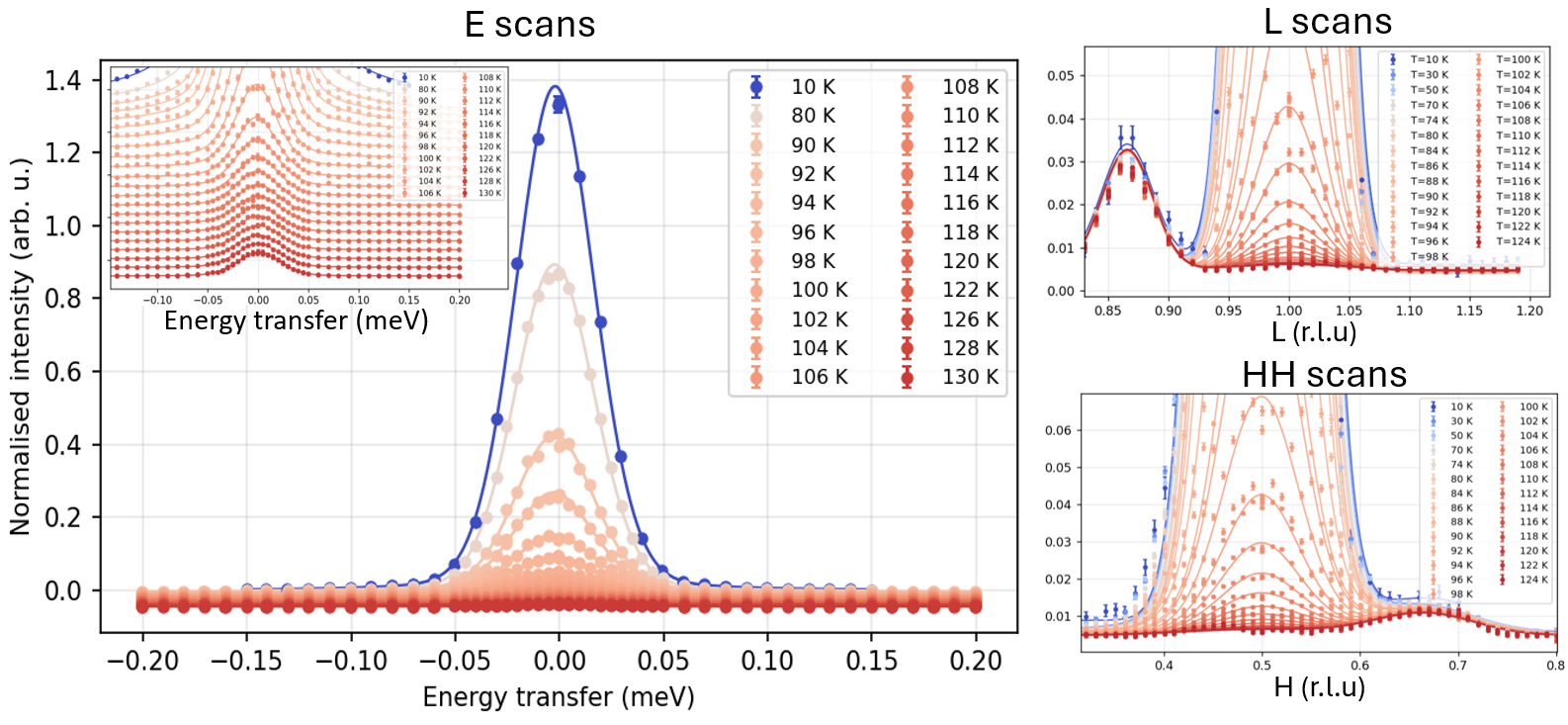}
    \caption{
      \textbf{Temperature evolution of the magnetic Bragg peak $(0.5,0.5,1)$.}
Normalized intensity scans are shown as a function of energy transfer $E$ (left), and as momentum scans along $L$ (top right) and along $(H,H)$ (HH, bottom right), for temperatures from \textbf{10~K to 130~K} (colour scale from low to high $T$). Experimental data are shown as symbols with error bars, and the corresponding fits are shown as solid lines. The inset in the $E$-scan panel shows a \textbf{zoom-in on the weak-intensity spectra}; for clarity, \textbf{each spectrum is vertically shifted by an arbitrary constant offset}.}
    \label{SM_fig:Mag}
\end{figure}

Figure \ref{SM_fig:Mag} summarizes neutron-scattering cuts acquired in the vicinity of the magnetic wave vector $\mathbf{Q}_{\mathrm{mag}}\simeq(0.5,0.5,1)$ with the sample mounted in the cryostat as described in the main text.
The background-subtracted $\mathbf{Q}$ and energy scans were analyzed using a pseudo-Voigt profile to parameterize the peak line shape in a model-independent manner. The intensity was described as a linear combination of Lorentzian and Gaussian contributions sharing a common full width at half maximum (FWHM) $w$,
\begin{equation}
I(E)=A\Big[\mu\,L(E;E_0,w)+(1-\mu)\,G(E;E_0,w)\Big]+I_{\mathrm{off}},
\end{equation}
where $A$ is the peak amplitude, $E_0$ the peak center, $I_{\mathrm{off}}$ a constant residual baseline, and $\mu$ the Lorentzian mixing fraction. The Lorentzian and Gaussian components were defined such that they are normalized to unity at $E=E_0$ and reach half maximum at $|E-E_0|=w/2$, so that $w$ corresponds directly to the peak FWHM in energy. 
Across the investigated temperature range, the fitted mixing coefficient was found to vary within $\mu\simeq 0.12$--$0.22$, indicating a predominantly Gaussian line shape with a finite Lorentzian component.

 In the $L$ scans an additional temperature-independent peak at $L\simeq0.88$~r.l.u.\ is attributed to an impurity/spurious contribution and is explicitly accounted for in the fits (via an extra Gaussian term) to avoid biasing the magnetic peak near $L\simeq1$.  
 
 Fewer data points were collected between 10 and 80~K, where no evolution was expected and is confirmed experimentally, and was made denser (typical step $\sim2$~K) in the temperature window of interest around the transition (between 80 K and 130 K),  above $\sim130$~K the magnetic signal becomes comparable to residual noise such that the relative uncertainty on the fitted amplitude exceeds 100\%, and these points are therefore shown with their error bars but are not used for reliable extraction of widths/FWHM. To note that the FWHM extracted along HH and $L$ at low temperature around $10$ K are the same (around 0.05 (r.l.u)) to the FWHM of selected nuclear Bragg peaks measured under identical instrumental conditions.

\bibliography{ref}